\documentclass{JHEP3}
\usepackage{amsmath}
\usepackage{graphicx}
\usepackage{float}

\title{MHV Diagrams from an All-Loop Recursion Relation}
\author{Mathew Bullimore\\ 
			Rudolf Peierls Centre for Theoretical Physics,\\
	 		1 Keble Road, Oxford, OX1 3NP,\\
			United Kingdom\\
			E-mail: \email{m.bullimore1@physics.ox.ac.uk}}

\abstract{We consider the recursion relation for loop integrands in planar $\cN=4$ SYM generated by an all-line shift of momentum twistors. We examine the behaviour of the rational loop integrands when the shift parameter becomes large, and find that a valid recursion relation may be obtained in all cases. The recursion relation is then formulated both in region momentum space and in momentum twistor space, and solved in detail for some one and two-loop examples. Finally, we show that the general iterative solution of the recursion relation generates the MHV vertex expansion for all loop integrands, providing a proof of the MHV diagram formalism for all loop amplitudes in planar $\cN=4$ SYM. }

\def \be  {\begin{equation}}
\def \ee  {\end{equation}}
\def \ba  {\begin{eqnarray}}
\def \ea  {\end{eqnarray}}
\def \lab #1 {\label{#1}}
\def \ni {\noindent}
\def \nn {\nonumber}

\newcommand\cA{\mathcal{A}}

\newcommand\cO{\mathcal{O}}

\newcommand\cN{\mathcal{N}}

\newcommand\CP {\mathbb{CP}}

\newcommand\rd{\mathrm{d}}
\newcommand\rD{\mathrm{D}}
\newcommand\la{\langle}
\newcommand\ra{\rangle}

\newcommand\dal{\dot{\alpha}}

\newcommand{\rf}{*}
\newcommand\im{i\!-\!1\,i}
\newcommand\jm{j\!-\!1\,j}
\newcommand\km{k\!-\!1\,k}
\newcommand\Nk{\mathrm{N}^k\mathrm{MHV}}

\newcommand\lb{\lambda}
\newcommand\tlb{\tilde{\lambda}}
\newcommand\al{\alpha}

\newcommand\Nsq{\mathrm{N}^2\mathrm{MHV}}

\newcommand\GL{\mathrm{GL}}

\begin{document}

\section*{Introduction}
\label{sec:Intro}

There have been rapid developments in the calculation of scattering amplitudes in gauge theory in recent years, largely inspired by twistor string theory~\cite{Witten:2003nn}. An early success was the development of the MHV diagram expansion for tree-level amplitudes~\cite{Cachazo:2004kj} motivated by the disconnected formulation of twistor string theory. The MHV diagrams are Feynman-like diagrams where the vertices are MHV amplitudes, continued off-shell by introducing an auxiliary  reference spinor, and the propagators are `$1/p^2$' Feynman propagators. The MHV diagram formalism offers substantial simplifications on the standard Feynman rules and leads to very compact expressions for tree amplitudes.

Since the original discovery, the MHV diagram formalism has been extended to gauge theories with scalar and fermion external states both massless~\cite{Georgiou:2004wu,Georgiou:2004by} and massive~\cite{Boels:2007pj,Boels:2008du} and have been used to compute one-loop amplitudes in supersymmetric gauge theories~\cite{Brandhuber:2004yw,Bedford:2004py,Quigley:2004pw}.  The lagrangian origin of the MHV diagrams has also been exposed both in spacetime~\cite{Mansfield:2005yd,Ettle:2006bw} and in twistor space~\cite{Mason:2005zm,Boels:2006ir,Boels:2007qn}. In maximally supersymmetric gauge theory, the MHV diagram formalism for has recently been formulated in a dual superconformally invariant manner for all loop amplitudes~\cite{Bullimore:2010pj} and shown to arise from the perturbative expansion of a supersymmetric Wilson loop in momentum twistor space~\cite{Mason:2010yk}.

Another important development in the computation of scattering amplitudes has been the discovery of BCFW recursion relations~\cite{Britto:2004ap,Britto:2005fq} which lead to compact expressions for tree-level scattering amplitudes. Inspired by BCFW, Risager introduced a recursion relation for gluon amplitudes by shifting the $(k+2)$ anti-holomorphic spinors $\tlb_i$ of negative helicity gluons and showed that iterative use of the recursion relations leads to the MHV expansion for tree amplitudes of gluons~\cite{Risager:2005vk}. The authors of~\cite{Elvang:2008vz} extended the Risager recursion and considered an all-line shift of anti-holomorphic spinors
\begin{equation}
\tlb^{\dal}_i \longrightarrow \tlb^{\dal}_1 + zc_i\, \zeta^{\dal} \qquad i=1,\ldots, n
\end{equation}
where $\zeta^{\dal}$ is an auxiliary reference spinor, and the coefficients $c_i$ are chosen to ensure momentum conservation. The iterative solution of the all-line recursion relation gives the MHV expansion for all tree amplitudes in maximally supersymmetric gauge theory. Recently, the all-line recursion relation has been extended to generic field theories and the conditions for the solution to generate the MHV expansion derived~\cite{Cohen:2010mi}. The purpose of this paper is to extend the all-line recursion relations to loop amplitudes in supersymmetric gauge theories.

The BCFW recursion relations have been extended to the rational parts of one-loop amplitudes~\cite{Bern:2005hs,Bern:2005cq} in non-supersymmetric gauge theories, and recently, to the rational integrands of all loop amplitudes in maximally supersymmetric gauge theory~\cite{ArkaniHamed:2010kv} (see also~\cite{Boels:2010nw}) generating compact expressions and revealing an exact Yangian symmetry~\cite{Drummond:2009fd} of the loop integrands to all loop orders. An important role in defining the loop integrand in planar gauge theories is played by the region coordinates $p_i^{\al\dal} = x^{\al\dal} _{i+1}-x^{\al\dal} _i$, which form the cusps of the null polygon in region momentum space.

\begin{figure}[htp]
\centering
\includegraphics[width=11cm]{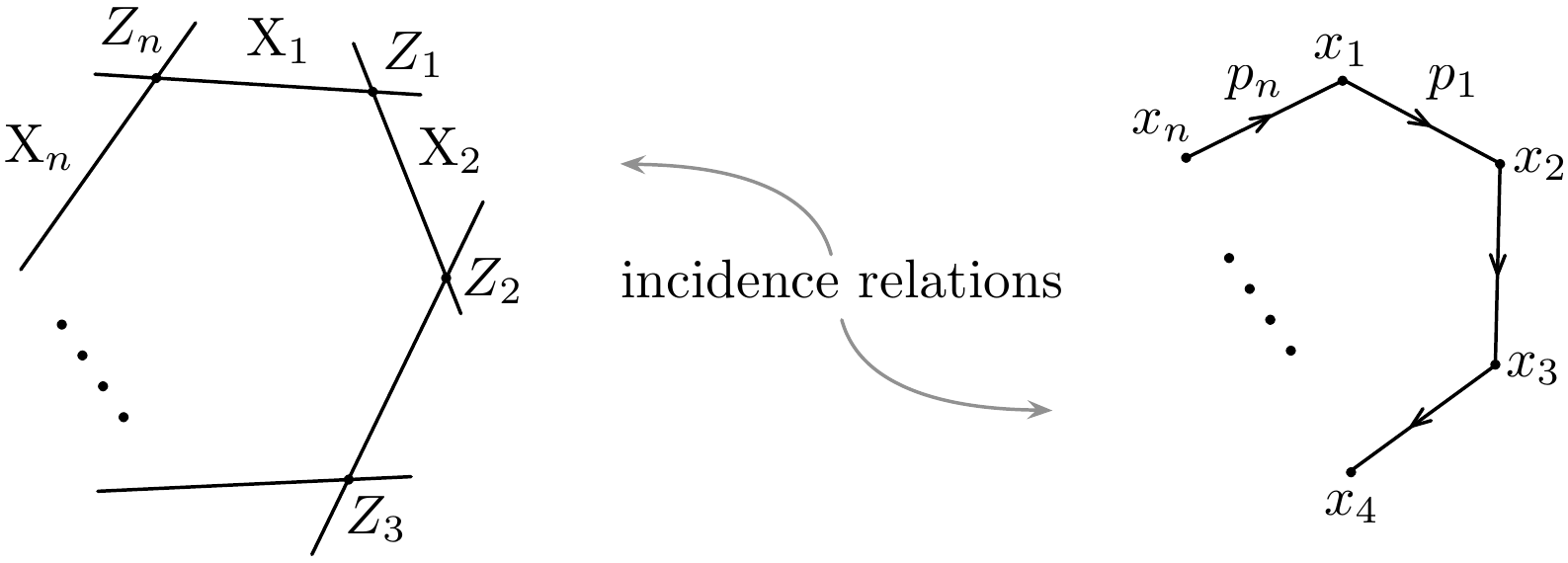}
\end{figure}

\ni An equivalent description of the null polygon is through the momentum twistor correspondence~\cite{Hodges:2009hk,Mason:2009qx}, where any set of momentum twistors $Z_i$  with components $(\lb_{i\al},\mu_i^{\dal})$ defines a null polygon in the region momentum space with cusps
\begin{equation}
x_i^{\al\dal} = \frac{\lb_{i-1}^{\al}\mu_i^{\dal} - \lb_i^{\al}\mu_{i-1}^{\dal}}{\la i-1\,i\ra}\, .
\end{equation}
The loop integrand is then a rational function of the momentum twistors $Z_i$ and the lines $(AB)_m$ in momentum twistor space where $m=1,\ldots,\ell$ and $\ell$ is the number of loops. Many examples of momentum twistor integrands with one and two loops have now been directly integrated on physical contours~\cite{Hodges:2010kq,Mason:2010pg,Drummond:2010mb,Alday:2010jz}.

We will consider the recursion relation for the loop integrand generated by the complex shift of the $\mu$-components of all the external momentum twistors
\begin{equation}
\mu^{\dal}_i \rightarrow \mu^{\dal}_i + z r_i \zeta^{\dal}
\end{equation}
\ni where $r_i$ are arbitrary coefficients and $\zeta^{\dal}$ is the auxiliary reference spinor. For tree-level amplitudes, this is equivalent to the all-line shift of anti-holomorphic spinors, and the amplitudes fall away as $\mathcal{O}(z^{-k})$ for $\Nk$ amplitudes in agreement with~\cite{Elvang:2008vz}. The loop integrands in maximally supersymmetric gauge theory behave as $\cO(z^{-k-\ell-1})$
where $\ell$ is the number of loops and hence valid all-line recursion relations may be obtained. The recursion relation may be represented schematically by 
\begin{figure}[htp]
\centering
\includegraphics[width=15.5cm]{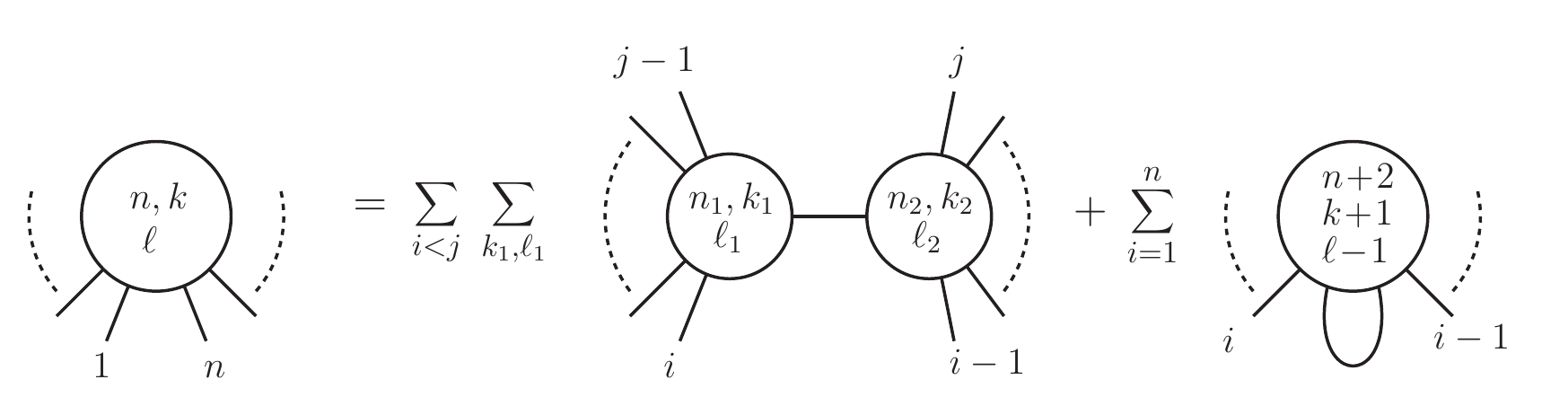}
\end{figure}

The all-line recursion relation shares many features in common with the BCFW recursion relation~\cite{ArkaniHamed:2010kv}, involving factorisation channels of the kind that already appear at tree-level, and also terms involving forward limits of integrands with one less loop. When the same reference spinor $\zeta^{\dal}$ is chosen at every stage, we will show that the iterative solution of the recursion relation is exactly the MHV diagram expansion for all loop amplitudes, providing a proof of the MHV formalism in $\cN=4$ SYM. 

The paper is organised as follows. In section~\ref{sec:MHVandIntegrand} we review the MHV diagram formalism for loop integrands in momentum space and momentum twistor space. In section~\ref{sec:TheShift} we introduce the all-line shift of momentum twistor and derive the $z\rightarrow\infty$ behaviour of loop integrands. The all-line recursion relations are then examined both in momentum space and momentum twistor space in section~\ref{sec:Recursion}. The recursion relations are then solved for some tree-amplitudes in section~\ref{sec:TreeLevel} and for one-loop and two-loop examples in section~\ref{sec:LoopLevel}. In section~\ref{sec:Proof} we prove the MHV formalism for all loop amplitudes in $\cN=4$ supersymmetric gauge theory using the all-line recursion relation. Finally, we present our conclusions and directions for future research in~\ref{sec:Conclusions}.


\section{MHV Rules and the Loop Integrand}
\label{sec:MHVandIntegrand}

In this section we review the supersymmetric version of the MHV diagram formalism appropriate for computing the rational integrands of superamplitudes in $\cN=4$ SYM. We will present the formalism in the standard (region) momentum space form and then the momentum twistor space formulation. However we must first define what is meant by the loop integrand.


\subsection{The Loop Integrand}

The integrands of planar loop amplitudes are rational functions of the external and internal loop momenta. The loop integrands are naturally chiral and possess all of the symmetries of the theory, which may be broken only by divergences on performing the integrals.  In maximally supersymmetric gauge theory, for example, the authors of~\cite{ArkaniHamed:2010kv} have shown that the loop integrand has an exact Yangian symmetry~\cite{Drummond:2009fd} combining the standard superconformal invariance with an additional dual superconformal symmetry~\cite{Drummond:2008vq}. The standard superconformal invariance has previously been shown in under the assumption of the MHV formalism for all loop amplitudes~\cite{Sever:2009aa}.

In general field theories, the loop integrand is not well-defined, since there is a translational freedom in the choice of origin for loop momenta. However, for planar amplitudes, where there is a well-defined ordering of the external particles, we may define region momenta $\{x_i\}$ as the cusps of the null polygon 
\begin{equation}
p^{\al\dal}_i = x_{i+1}^{\al\dal} - x_i^{\al\dal}
\end{equation}
Region momenta $\{y_m\}$ may also be assigned to internal loops, absorbing the translational freedom, and allowing a sharp definition of the loop integrand. For superamplitudes in $\cN=4$ SYM with $n$ particles, $\ell$ loops, and of degree $\Nk$, the integrand is defined by 

\begin{equation}
\label{Integrand1}
\cA_{n,k}^{(\ell)}(1,\ldots,n) = \int \prod\limits_{m=1}^{\ell} \rd^4 y_m\;  A_{n,k}^{(\ell)}(1,\dots,n;y_m)
\end{equation}

\ni together with the prescription that the integrand is symmetrised over the assignment of region coordinates $y^{\al\dal}_m$ to internal loops ($m=1,\ldots,\ell$). The amplitude and the integrand then have overall grassmann degree $(8+4k)$. 

An equivalent solution is to introduce momentum twistors $Z_i$ which automatically define a null polygon in region momentum space. Internal regions become lines in momentum twistor space and are represented by two momentum twistor coordinates $A_m$ and $B_m$ modulo $GL(2)$ transformations that send $A_m$ and $B_m$ along the line $(AB)_m$. The momentum twistor integrand is then defined by the equation

\begin{equation}
\label{Integrand2}
A_{n,k}^{(\ell)}(Z_1,\ldots,Z_n) = \int\prod\limits_{m=1}^{\ell}\left[ \frac{\rd^{4|4}A_m \rd^{4|4}B_m}{\mathrm{vol\,GL}(2)} \right] \; I_{n,k}^{(\ell)}(Z_1,\ldots,Z_n,(AB)_m)
\end{equation}
\pagebreak

\ni where again the integrand is symmetrised over the lines $(AB)_m$ in momentum twistor space. Compared to equation~\eqref{Integrand1} the momentum twistor integrand has an overall MHV amplitude amputated and, in addition, the fermionic integrals $\prod_m \rd^4\chi_{A_m} \rd^4\chi_{B_m}$ have been pulled out of the momentum twistor integrand and included in the measure. This will allow dual superconformally invariant expressions for the loop integrands which now have grassmann degree $4(k+2\ell)$.


\subsection{MHV Diagrams in Momentum Space}

Let us first consider MHV diagrams for the integrands of superamplitudes in $\cN=4$ SYM, formulated in terms of region momentum space. The vertices are formed from the MHV superamplitude
\begin{equation}
A^{(0)}_{\mathrm{MHV}}(1,\ldots,n) = \frac{\delta^{0|8}\left(\sum\limits_{i=1}^n \lb_i \eta_i\right)}{\la 12\ra\ldots\la n1\ra}\, .
\end{equation}

\ni When the vertex is connected to a propagator bounding regions $x^{\mu}$ and $y^{\mu}$ which may be internal or external, then the following spinor is assigned to the propagator

\begin{equation}
\lb^\al = (x-y)^{\al\dal}\zeta_{\dal} 
\end{equation}

\ni where $\zeta^{\dal}$ is a reference spinor whose dependence drops out in summing over all diagrams. In addition, for each propagator there is a grassmann integration $\rd^4\eta$ which sums over all helicity states in the supermultiplet. The MHV diagrams for the loop integrand that differ by an exchange of internal region momenta are considered as separate diagrams. The sum of diagrams then comes with an overall factor $1/\ell\,!$ associated with the symmetrisation condition.


\subsection{MHV Diagrams in Momentum Twistor Space}

The MHV diagram formalism has been translated into momentum twistor space~\cite{Bullimore:2010pj} where the formalism is dual superconformally invariant modulo the choice of a reference twistor. The MHV rules have also recently been written in dual superspace in~\cite{Brandhuber:2010mi}. In this setting, the MHV diagram formalism arises naturally from the perturbative expansion of a supersymmetric Wilson loop in momentum twistor space using the twistor action in the axial gauge~\cite{Mason:2010yk}.

In momentum twistor space the vertices become one and the propagators are associated with dual superconformal invariants $[\,*\,,\ ,\ ,\ ,]$ where $Z_*$ is an arbitrary reference twistor. The fundamental dual superconformal invariant is defined by the following expression

\begin{equation}
\label{Invariantdef}
[a,b,c,d,e]  = \frac{\delta^{0|4}\left(\, \la a\,b\,c\,d \ra\eta_e + \mathrm{cyclic}\, \right)}{\la a\,b\,c\,d \ra\la b\,c\,d\,e \ra\la c\,d\,e\,a \ra\la d\,e\,a\,b \ra\la e\,a\,b\,c \ra}\, .
\end{equation}

\ni and is essentially a supersymmetric delta-function ensuring that five points are linearly dependent in super-twistor space. \pagebreak They are related to the more familiar `R-invariant' expressions by $R_{n;ij} = [n,i\!-\!1,i,j\!-\!1,j]$. Such invariants arise whenever there is a factorisation channel in momentum space. The standard momentum space rules are recovered when the reference twistor takes the particular form $Z_* = (0,\zeta^{\dal},0)$ upon which $\zeta^{\dal}$ becomes the standard reference spinor.

Let us now summarise the MHV rules for planar diagrams in $\cN=4$ SYM with $\ell$ loops in momentum twistor space~\cite{Bullimore:2010pj}:

\begin{itemize}

	\item External regions $x_i$ correspond to lines $X_i$ in momentum twistor space, which pass through pairs 		of momentum twistors $Z_{i-1}$ and $Z_i$.
	
	\item Internal regions are associated with lines $(AB)_m$ in momentum twistor space, which pass through two 		momentum twistors $A_m$ and $B_m$ (with $m=1,\ldots,\ell$).

	\item To every propagator in the planar diagram, assign an invariant $[\,\rf\,,\ ,\ ,\ ,\ ]$ depending on the reference twistor $Z_\rf$ and two pairs of momentum twistors corresponding to the two regions (external or internal) that are bounded by the propagator.
\end{itemize}

The general rule for assignment of momentum twistors depends on choosing an orientation for the diagrams, and we choose the anti-clockwise orientation around each vertex.  Suppose that the regions $u,v,\ldots$, which may be external or internal, are associated with the lines $(U_1 U_2),(V_1 V_2),\ldots$ in momentum twistor space. The general assignment of momentum twistors may then be neatly summarised by the following\footnote{The assignment of momentum twistors here differs from that described~\cite{Bullimore:2010pj} to have identical assignments for both internal and external regions. However, the two sets of assignments are equivalent.}:
\begin{figure}[htp]
\centering
\includegraphics[width=11.5cm]{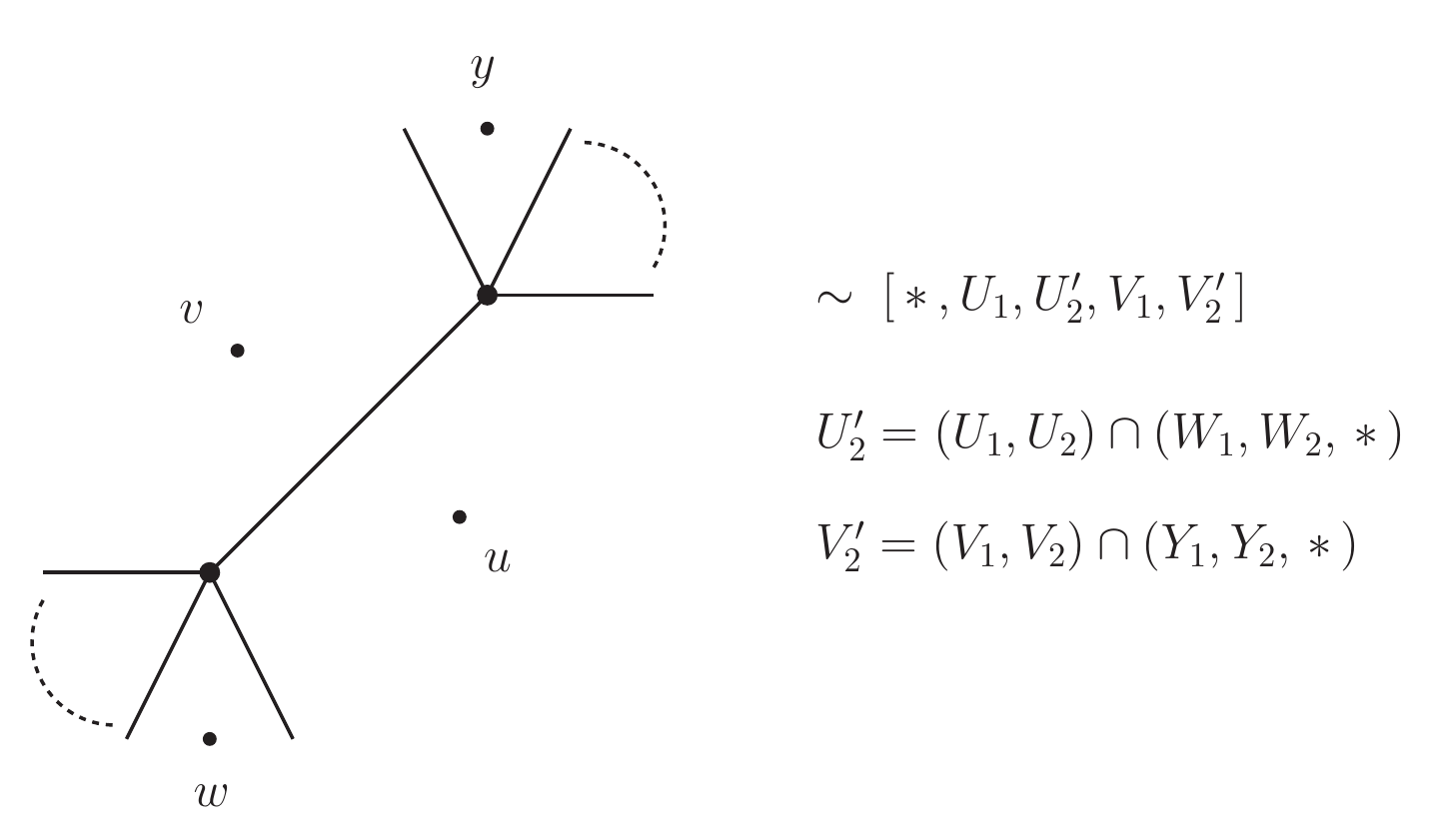}
\end{figure}
	
\begin{figure}[htp]
\centering
\includegraphics[width=14cm]{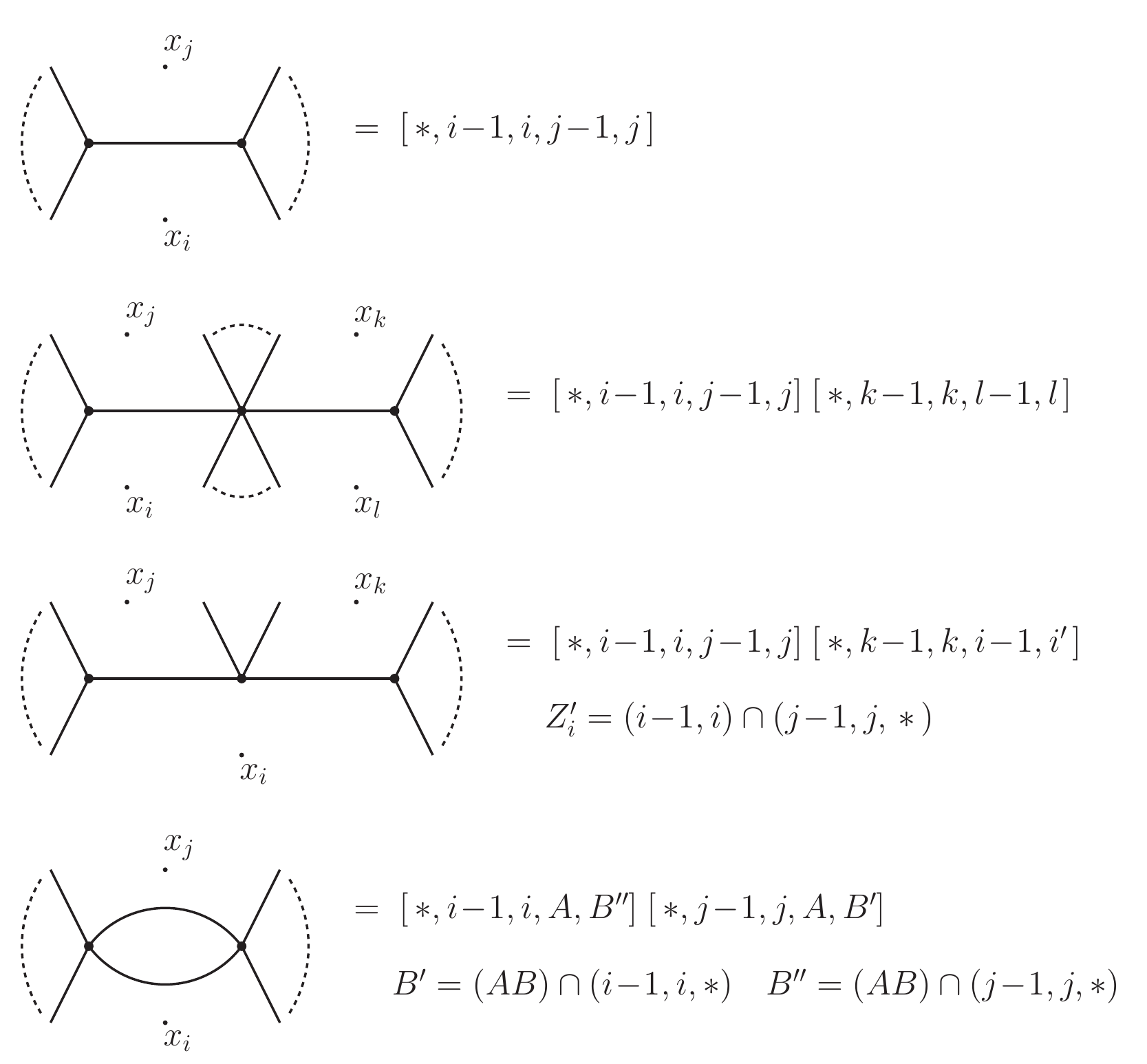}
\caption{\emph{Examples to illustrate evaluation of MHV diagrams in momentum twistor space.}}
\label{fig:MHVrulesexamples}
\end{figure}

\ni For example, the MHV diagrams for the tree-level NMHV amplitude have a single channel $p=(x_i-x_j)$ bounding two external regions $x_i$ and $x_j$, and correspond to the single invariant $[\,*\,,i\!-\!1,i,j\!-\!1,j]$. Some further simple examples examples of the momentum twistor MHV diagrams are illustrated in figure~\ref{fig:MHVrulesexamples}.

\section{The All-Line Shift}
\label{sec:TheShift}

In this section we introduce the all-line shift for loop integrands. The shift is described in both momentum twistor space and as a shift of all external region momenta. We also show that loop integrands in $\cN=4$ SYM fall off as $\mathcal{O}(z^{-k-\ell-1})$ as $z\rightarrow\infty$ under the all-line shift.


\subsection{Momentum Twistor Shift}

Consider the following complex shift of all momentum twistors

\begin{equation}
\label{Momtwistorshift}
Z_i \rightarrow Z_i + z r_i Z_* \qquad i = 1,\ldots,n\,
\end{equation}

\ni where $r_i$ are non-zero complex coefficients and the reference twistor has only non-zero secondary component $Z_*=(0,\zeta^{\dal},0)$. Any choice of $n$ momentum twistors automatically defines $n$ four-momenta forming a closed null polygon, and therefore momentum conservation is guaranteed.

The momentum twistors have components $Z = (\lb_\al,\mu^{\dal},\eta^a)$ and therefore the complex shift effects only the secondary part of the momentum twistors

\begin{equation}
\mu_i^{\dal} \rightarrow \mu_i^{\dal} + z r_i \zeta^{\dot{\al}}
\end{equation}

\ni The momentum twistors define a null polygon whose cusps are defined by

\begin{equation}
x_i^{\al\dal} = \frac{\lb_{i-1}^{\al}\mu_i^{\dal} - \lb_i^{\al} \mu_{i-1}^{\dal}}{\la i-1\,i\ra} \, .
\end{equation}

\ni with a similar equation for the fermionic coordinates. Hence we may work out the corresponding shifts of the region momenta as follows

\begin{eqnarray}
\label{RegionShift}
x_i &\rightarrow& x_i + z \left[ \frac{\lb_{i-1}r_i - \lb_i r_{i-1}}{\la i-1\, i \ra} \right] \zeta \nn\\
&\equiv& x_i + z q_i \zeta \, .
\end{eqnarray}

\ni The main motivation for choosing the momentum twistor shift~\eqref{Momtwistorshift} is the simple shift of the region momenta. The origin the spinors $q^\al_i$ is unimportant and the final solution will be independent of them. The key requirement is that all external region momenta are shifted in order to capture all factorisation channels in the loop integrand. 


\subsection{Shift of Anti-holomorphic Spinors}

We have seen that the left-handed spinors are unchanged by the shift. Let us now calculate the effect of the momentum twistor shift~\eqref{Momtwistorshift} on the right-handed spinors $\tlb$. The right-handed spinors are defined from the region momenta by

\begin{equation}
\tlb^{\dal}_i = \frac{(x_{i+1} - x_i)^{\al\dal}\lb_{i+1\,\al}}{\la i\, i+1\ra} \, .
\end{equation}

\ni and it is straightforward to show from equation~\eqref{RegionShift} that 

\begin{equation}
\label{All-lineShift}
\tlb_i \longrightarrow \tlb_i + z c_i \zeta \qquad i=1,\ldots,n
\end{equation}

\ni where the coefficients are

\begin{equation}
\label{Coefficients}
c_i = \frac{\la i-1\,i\ra r_{i+1} + \la i+1\,i-1\ra r_i + \la i\, i+1\ra r_{i-1}}{\la i-1\,i\ra\la i\,i+1 \ra}\, .
\end{equation}

\ni Hence we have the all-line shift introduced in~\cite{Elvang:2008vz} where the resulting recursion relation was solved and shown to generate the MHV diagram formalism for tree-amplitudes in $\cN=4$ SYM. However, when discussing the integrands of loop amplitudes, it is important to define the shift in terms of region momenta or momentum twistors. 


\subsection{Behaviour at Infinity}

In order to generate a useful recursion relation, there must be no pole in the integrand as the shift parameter tends to infinity $z\rightarrow\infty$. In this limit, all external momentum twistors are all sent towards the reference twistor $Z_*$ (projectively) corresponding to a multiple collinear limit in momentum space. Nevertheless, we will use the dual superconformal invariance of the loop integrand to show that an $\ell$-loop $\Nk$ integrand behaves as $\mathcal{O}(z^{-k-\ell-1})$ for $\ell\geq0$ and that the tree-level superamplitudes behave as $\mathcal{O}(z^{-k})$ in the limit $z\rightarrow\infty$.

\subsubsection*{Tree Amplitudes}

Consider first the tree-level superamplitudes. The BCFW expansion expresses the tree-level superamplitude as a sum of terms, each of which is the product of dual superconformal invariants $[\, ,\, ,\, ,\, ,\,]$ which may have shifted arguments~\cite{Drummond:2008cr}. Since the four bracket $\la\, ,\, ,\, ,\,\ra$ is completely antisymmetric in its arguments then for large-$z$ we have

\begin{equation}
\la\, ,\, ,\, ,\,\ra \sim \cO(z)\, .
\end{equation}

\ni The fermionic delta-function in each invariant $[\, ,\, ,\, ,\, ,]$ then behaves as $\mathcal{O}(z^4)$ and hence from the definition~\eqref{Invariantdef} we have

\begin{equation}
[\, ,\, ,\, ,\, ,] \sim \mathcal{O}(z^{-1})\, .
\end{equation}

\ni Each term in the BCFW expansion of tree-level superamplitudes is a product of $k$ invariants $[\, ,\, ,\, ,\, ,] $ possibly with shifted arguments. Hence the tree-level superamplitudes behave as $\mathcal{O}(z^{-k})$ in agreement with the results found in~\cite{Elvang:2008vz,Cohen:2010mi}.

\subsubsection*{Loop Integrands}

Now consider the loop integrands. For this subsection only we include the fermionic integration $\int \prod_m \rd^4\eta_{A_m}\rd^4\eta_{B_m}$ in the definition of the momentum twistor integrand, and hence loop integrands of N$^k$MHV superamplitudes then have grassmann degree $4k$. Following~\cite{ArkaniHamed:2010kv} we may expand the integrand in a basis of chiral integrands having unit leading singularities and hence zero grassmann degree. The coefficients in the expansion are then residues of the $G(k,n)$ grassmannian formula with grassmann degree $4k$. The grassmannian residues are products of $k$ fundamental dual superconformal invariants $[\, ,\, ,\, ,\, ,\,]$~\cite{Bullimore:2010pa} and hence behave as $\cO(z^{-k})$ as shown above. 

Now consider the chiral integrands with unit leading singularity and zero grassmann degree. Dual conformal invariance requires that they are constructed from four-brackets $\la\, ,\, ,\, ,\,\ra$ and have zero weight in the external momentum twistors $Z_i$ and weight $-4$ is the loop momentum twistors $A_m$ and $B_m$. Hence, any such one-loop integrand may be expressed as follows~\cite{ArkaniHamed:2010kv}

\begin{equation}
\frac{\la AB\,Y_1\ra \ldots \la AB\, Y_{n-4}\ra }{\la AB12 \ra\la AB23\ra \ldots\ldots \la ABn1 \ra}
\end{equation}

\begin{figure}[htp]
\centering
\includegraphics[width=10cm]{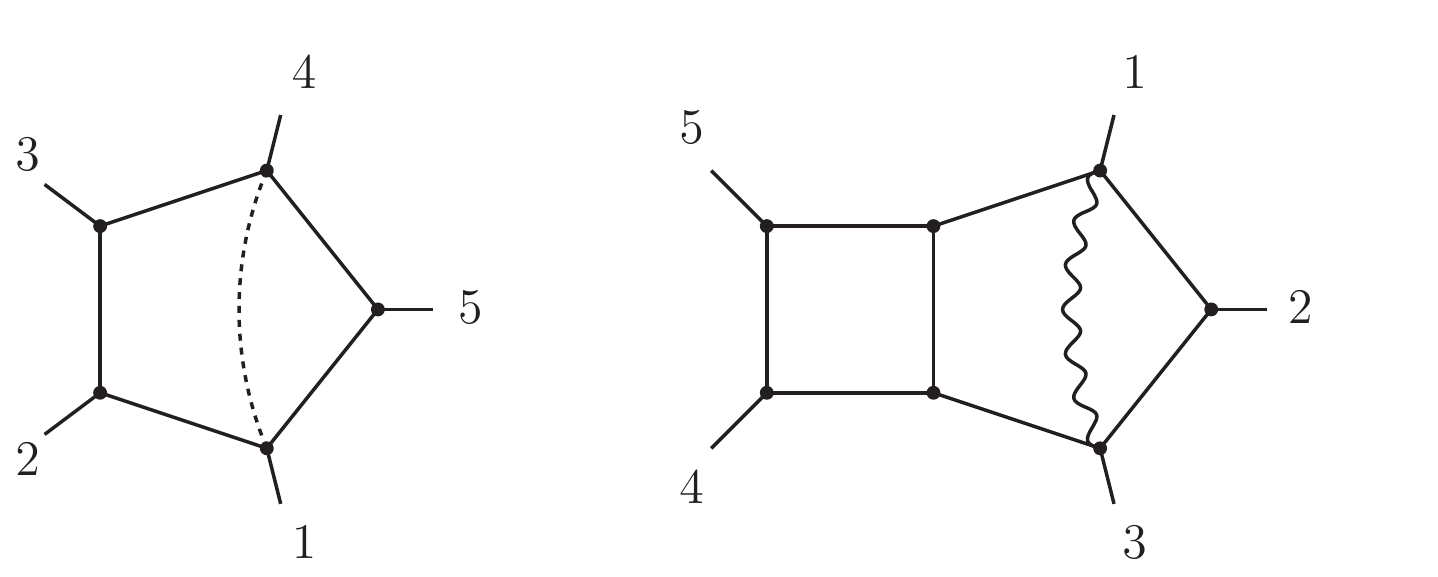}
\caption{\emph{Chiral integrals with unit leading singularities used to illustrate large $z$ behavior.}}
\label{fig:ChiralIntegrals}
\end{figure}

\ni where we have included all physical propagators in the denominator and where $Y_{1},\ldots,Y_{n-4}$ are antisymmetric twistors. The antisymmetric twistors $Y_1,\ldots,Y_{n-4}$ must together carry weight $2n$ in each external momentum twistor. Hence expanding each one in a basis of simple bitwistors then the coefficients must contain two four brackets $\la\, ,\, ,\, ,\,\ra$. For example, we have the chiral pentagon integrand (see figure~\ref{fig:ChiralIntegrals})

\begin{equation}
\frac{\la AB14 \ra\la 5123 \ra\la 2345 \ra}{\la AB12 \ra\la AB23 \ra\la AB34 \ra\la AB45 \ra\la AB51 \ra}\, .
\end{equation}

The four brackets corresponding to propagators in momentum space behave as

\begin{equation}
\la AB,\, ,\,\ra \sim \mathcal{O}(z)
\end{equation}

\ni and therefore, counting weights, the one-loop integrands of unit leading singularity behave as $\cO(z^{-2})$. Now combining this with the behaviour of the coefficients, each term in a local expansion has behavior $\mathcal{O}(z^{-k-2})$ and the same property is inherited by the complete one-loop integrand.

Since both kinds of four-bracket $\la\, ,\, ,\, ,\,\ra$ and $\la AB,\, ,\, \ra$ behave as $\cO(z)$ then increasing the number of loops can only improve the behavior of integrand with $z\rightarrow\infty$. For example, consider the two-loop pentabox integrand (see figure~\ref{fig:ChiralIntegrals})

\begin{equation}
\frac{ \la 3451 \ra\la 4513\ra \la AB| (512)\cap(234) \ra}{\la AB51 \ra\la AB12 \ra\la AB23 \ra\la AB34 \ra\la ABCD \ra\la CD 34 \ra\la CD45 \ra\la CD51 \ra}\, .
\end{equation}

\ni Expanding the numerator 
\begin{equation}
\la AB| (512)\cap(234) \ra = \la A512 \ra\la B234 \ra - \la B512 \ra\la A234 \ra
\end{equation}

\ni then it is clear that the integrand falls away as $\cO(z^{-3})$. For the integrands of unit leading singularity, an extension of the above argument shows that they fall away as $\cO(z^{-\ell-1})$. Combining with the grassmannian residues then each term in a local expansion falls away as $\cO(z^{-k-\ell-1})$ and again the same behaviour is inherited by the full integrand. This shows that there are no poles at infinity for any loop integrands under the all-line momentum twistor shift.


\section{The Recursion Relation}
\label{sec:Recursion}

In this section we discuss the all-loop recursion relation arising from the all-line shift in the previous section. We first present the recursion relation in both dual momentum space and momentum twistor space and then discuss in detail each of the contributions separately.

\subsection{The Recursion Relations}

Let us first consider the momentum space loop integrand. The shifted integrand $ A_{n,k}^{(\ell)}(z)$ depends on the shift parameter and we consider the following contour integral which encloses all poles in the complex $z$-plane

\begin{equation}
\oint \frac{\rd z}{z}\; A_{n,k}^{(\ell)}(z)=0  
\end{equation}

\begin{figure}[htp]
\centering
\includegraphics[width=15.5cm]{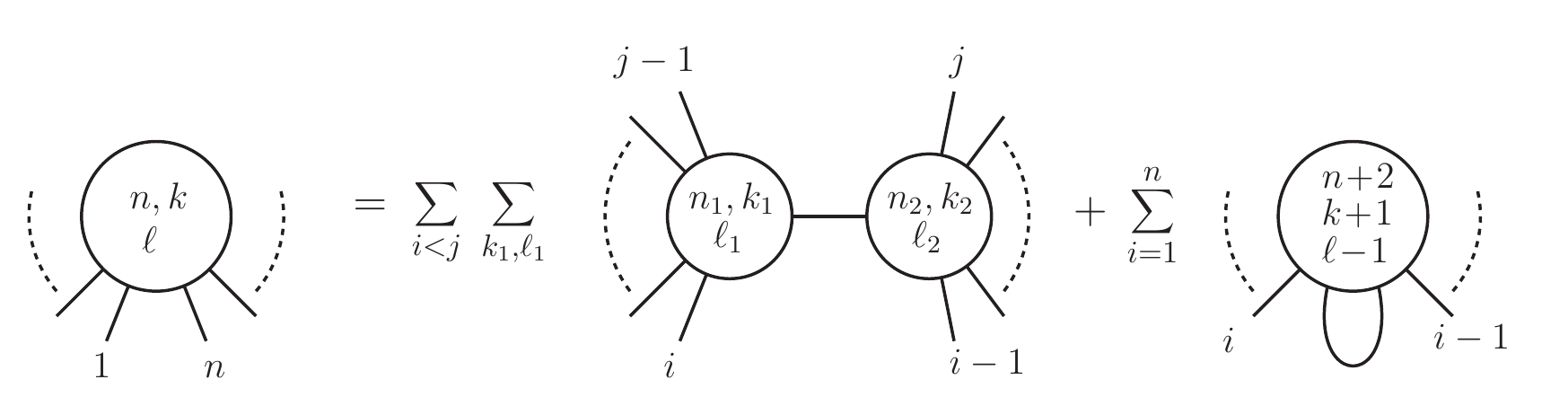}
\caption{\emph{A graphical representation of the all-line recursion relation for generic integrands.}}
\label{fig:RecursionSchematic}
\end{figure}

\ni and whose vanishing is ensured by the absence of any pole at infinity. The residue of the pole at $z=0$ is the original integrand and in addition there are poles arising from propagators which bound at least one external region. There are poles from propagators bounding two external regions, on which the integrand factorises into smaller integrands. However, there are also poles from propagators bounding an external and an internal region, which involve forward limits. The recursion relation is illustrated schematically in figure~\ref{fig:RecursionSchematic}.

The full recursion relations for the momentum space loop integrand are

\begin{eqnarray}
\label{RecursionRegion}
A_{n,k}^{(\ell)}(1,\ldots,n) &=& \sum\limits_{i,j,k_1,\ell_1}\; \int \rd^4\eta_I \; A_{n_1,k_1}^{(\ell_1)}(i,\ldots,j\!-\!1,I;z_I)\; \frac{1}{(x_i-x_j)^2}\; A_{n_2,k_2}^{(\ell_2)}(j,\ldots,i\!-\!1,-I;z_I) \nn\\
&+& \sum\limits_{i=1}^n \frac{1}{(x-x_i)^2}\int \rd^4\eta_I \; A_{n+2,k+1}^{(\ell-1)}(i,\ldots,i\!-\!1,I,-I;z_I)\, .
\end{eqnarray}

\ni The channels going on shell are labelled with subscripts $I$ and the notation $A(\ldots;z_I)$ means that all external region momenta are all shifted and evaluated on the relevant pole $z_I$. Also the external legs denoted by $I$ and $-I$ correspond to the variables
\begin{equation}
I = \{\lb_I,\eta_I\} \qquad -I = \{-\lb_I,\eta_I\}
\end{equation}
where $\lb_I$ is the CSW spinor associated with the off-shell momentum $P_I$ flowing in the channel $I$. This will be discussed further in the following subsections, which deal with the two kinds of term in more detail.

The summation ranges in the first line of the recursion relation~\eqref{RecursionRegion} are as follows. Firstly, there is a sum over the numbers of particles on the integrands $1\leq i < j \leq n$ where $n_1+n_2=n+2$. In addition there is a summation over the grassmann degree $0\leq k_1\leq k-1$ where $k_2+k_2=k+1$, and number of loops $0\leq\ell_1\leq\ell$ where $\ell_1+\ell_2=\ell$. Although the internal regions have not been denoted explicitly, the whole expression must be symmetrised over the assignment of region momenta $y_m$ with $m=1,\ldots,\ell$ to internal regions. 

For tree-level superamplitudes, there are only standard factorisation terms, and hence we have the simpler all-line recursion relation
\begin{equation}
\label{RecursionRegionTree}
A_{n,k}^{(0)}(1,\ldots,n) = \sum\limits_{i,j,k_1}\; \int \rd^4\eta_I \; A_{n_1,k_1}^{(0)}(i,\ldots,j\!-\!1,I;z_I)\; \frac{1}{(x_i-x_j)^2}\; A_{n_2,k_2}^{(0)}(j,\ldots,i\!-\!1,-I;z_I) 
\end{equation}

\ni which has previously been studied in~\cite{Elvang:2008vz} and more recently in~\cite{Cohen:2010mi} in the context of general field theories.

We will also formulate the recursion relations directly in momentum twistor space employing the techniques developed in~\cite{ArkaniHamed:2010kv}. This formulation has the advantage that it generates the MHV diagrams rules in their dual superconformally invariant form. The full recursion relations for the momentum twistor integrand are 

\begin{eqnarray}
\label{RecursionTwistor}
A_{n,k}^{(\ell)}(1,\ldots,n) &=& \sum\limits_{i,j,k_1,\ell_1} [\,*\, ,i\!-\!1,i,j\!-\!1,j]\; A_{n_1,k_1}^{(\ell_1)}(Z_I,i,\ldots,j\!-\!1;z_I )\; A_{n_2,k_2}^{(\ell_2)}(Z_I,j,\ldots,i\!-\!1;z_I) \nn\\
&+&\sum\limits_{i=1}^{n}\; \oint\limits_{\mathrm{GL}(2)} [\,*\, ,i\!-\!1,i,A,B] \; A_{n+2,k+1}^{(\ell-1)}(i,\ldots,i\!-\!1,A,B';z_I) \, . 
\end{eqnarray}

\ni where the summations are the same as the preceding paragraph. The momentum twistor $B'$ is defined by the intersection $(AB)\cap(i-1,i,*)$ and the momentum twistor $Z_I$ is a natural point in the geometry described below. The external momentum twistors $Z_i$ are all shifted and evaluated on the relevant pole $z_I$, and for multi-loop amplitudes, the expression must be symmetrised over the possible lines $(AB)_m$.


\subsection{Factorisation Terms}

\begin{figure}[htp]
\centering
\includegraphics[width=6cm]{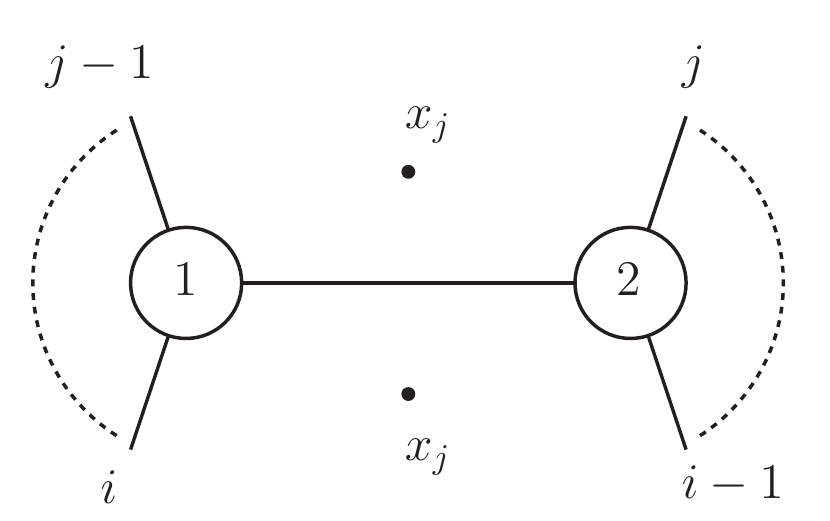}
\caption{\emph{The standard factorisation terms in the all-line recursion relation.}}
\label{fig:Standardfactorisation}
\end{figure}

Firstly, there are poles in the integrand from propagators which bound two external regions, which arise for both tree and loop amplitudes. Consider the pole arising from the propagator $(x_i-x_j)^2$ with momentum in the channel $I=\{i,\ldots,j-1\}$. The pole occurs when the shifted momentum $P_I(z) = x_{ij}(z)$ becomes null:
\begin{equation}
z_I = -\frac{x_{ij}^2}{2\la q_I |x_{ij}|\zeta]}
\end{equation}
where $q^{\al}_I = q^{\al}_i-q^{\al}_j$ and the coefficient spinors are defined in equation~\ref{RegionShift}. The standard LSZ arguments ensure factorisation of the integrand into integrands containing fewer numbers of particles and fewer numbers of loops - see figure~\ref{fig:Standardfactorisation}.

\subsubsection*{Region Momentum Space}

In region momentum space the contribution to the recursion relation is then

\begin{equation}
\int \rd^4\eta_I \; A_{n_1,k_1}^{(\ell_1)}(i,\ldots,j\!-\!1,I;z_I)\; \frac{1}{(x_i-x_j)^2}\; A_{n_2,k_2}^{(\ell_2)}(j,\ldots,i\!-\!1,-I;z_I)
\end{equation}

\ni where all external region momenta are shifted and evaluated on the pole $z_I$. The on-shell momentum $P_I(z_I)$ may be written $\lb_I\tlb_I$ and contracting with the reference spinor $\zeta^{\dal}$ we find that the holomorphic spinor associated with the off-shell propagator is $\lb_I = P_I\, |\zeta] / [ \tlb_I\,\zeta]$. However, the product of integrands together with the fermion measure $\rd^4\eta_I$ are invariant under the little group rescaling $(\lb_I,\eta_I)\rightarrow(t\lb_I,t^{-1}\eta_I)$ and hence we may equally use the holomorphic spinor 
\begin{equation}
\lb_I = P_I |\zeta] = (x_i - x_j) |\zeta]
\end{equation}

\ni which is the holomorphic CSW spinor associated with the off-shell propagator~\cite{Cachazo:2004kj}. The notation $I$ and $-I$ is then shorthand for the variables $\{ \lb_I,\eta_I\}$ and $\{ -\lb_I,\eta_I\}$. 
\subsubsection*{Momentum Twistor Space}

\begin{figure}[htp]
\centering
\includegraphics[width=6.5cm]{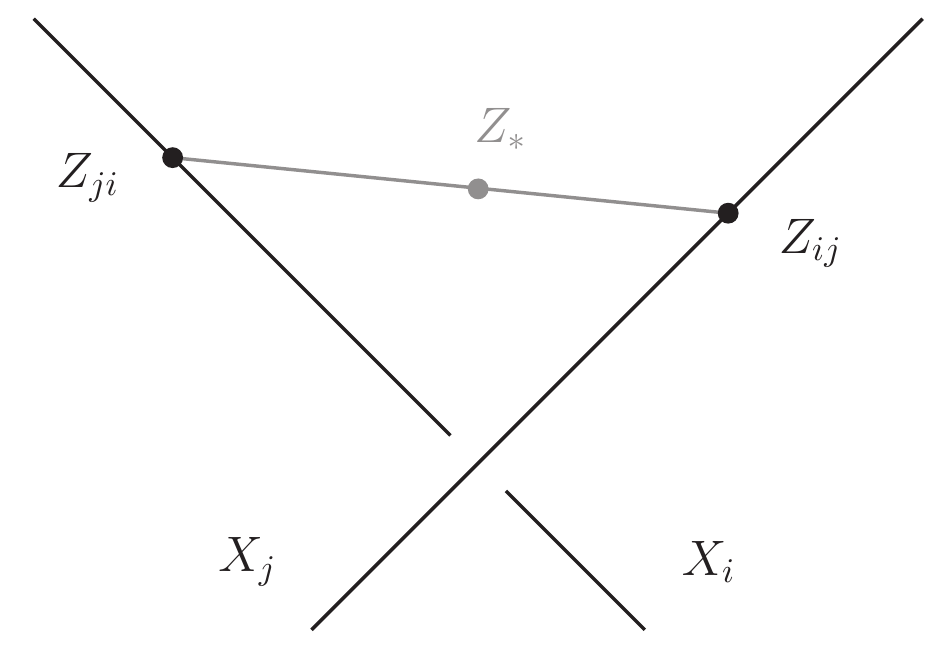}
\caption{\emph{The geometry of poles in the loop integrand in momentum twistor space.}}
\label{fig:Factorisationgeometry1}
\end{figure}

Let us now understand the momentum twistor geometry of the factorisation terms. First consider that there are no shifts of the momentum twistors. The two regions $x_i$ and $x_j$ correspond to lines $X_i$ and $X_j$ in momentum twistor space which are generically skew. However, the reference twistor determines a unique line intersecting $X_i$ and $X_j$ which then defines two distinguished points on those lines (see figure~\ref{fig:Factorisationgeometry1})

 \begin{equation}
Z_{ij} = \la \,*\,\im\, [\,j\!-\!1\ra\, Z_j]
\qquad
Z_{ji} = \la \,*\,\jm\, [\,i\!-\!1\ra\, Z_i]
\end{equation}

\ni which are the intersection points $(\im)\cap(\jm\,*)$ and $(\jm)\cap(\im\,*)$ in the geometry. The supersymmetric extension is valid on the support of the fermionic delta-functions in the fundamental invariant $[\,*\, ,i\!-\!1,i,j\!-\!1,j]$ which arises from the propagator.

\pagebreak
However, there is another distinguished point in the geometry since three points on a line automatically define a fourth $Z_I$ which is equi-anharmonic with the other three\footnote{I am indebted to Lionel Mason for bringing to my attention the importance of this point.}. This means that the four points have cross-ratio $-1$. The components of  $Z_I$ are 
\begin{eqnarray}
\label{WP1}
Z_I&=&Z_{ij} + \frac{1}{2}\la \im\,\jm \ra Z_* \nn\\
&=& Z_{ji}  + \frac{1}{2}\la \im\, \jm\ra Z_* \, .
\end{eqnarray}
\ni The equality of the bosonic components follows from the linear dependence of five bosonic momentum twistors and equality of the fermionic components follows on the support of the fermionic delta-functions in $[\,*\, , i\!-\!1,i,j\!-\!1,j\,]$. This special point in the geometry plays an important role in the recursion relation.

Now consider the shift of all external momentum twistors
\begin{equation}
Z_i \longrightarrow Z_i + zr_iZ_*
\end{equation}
\ni under which the shifted lines $X_i$ and $X_j$ are shifted in the planes $(\im\,*)$ and $(\jm\,*)$ respectively. The pole in the integrand then occurs when the lines $X_i(z)$ and $X_j(z)$ intersect and one may show (for example, by using incidence relations) that this happens at the equi-anharmonic point $Z_I$. The primary component of $Z_I$ is then exactly the holomorphic CSW spinor $\lb_I$ associated with the off-shell propagator.

The contribution from the residue at this pole is then
\begin{equation}
[\,*\, ,i\!-\!1,i,j\!-\!1,j]\; A_{n_1,k_1}^{(\ell_1)}(I,i,\ldots,j-1;z_I)\; A_{n_2,k_2}^{(\ell_2)}(I,j,\ldots,i-1;z_I)
\end{equation}
\ni where the external momentum twistors are evaluated on the parameter $z_I$ where the pole occurs. Note that the propagator contributes an invariant $[\,*\,,i\!-\!1,i,j\!-\!1,j]$. The contribution of an invariant with a propagators is  a direct consequence of translating factorisation channels into momentum twistor space~\cite{Bullimore:2010pj}.

\pagebreak
\subsection{Forward Terms}

There are also simple poles in the loop integrand from propagators bounding an internal and an external region. Consider the pole arising from the propagator $(x-x_i)^2$ where $x$ is an internal region, which occurs when the four-momentum $P_I(z)=(x-x_i(z))$ becomes null 
\begin{equation}
z_I = \frac{(x-x_i)^2}{2\la q_i|(x-x_i)|\zeta]}
\end{equation}
Standard arguments ensure factorisation and the residue then corresponds a forward limit in the channel $(x-x_i)$ of an integrand with one less loop. 

\begin{figure}[htp]
\centering
\includegraphics[width=6.5cm]{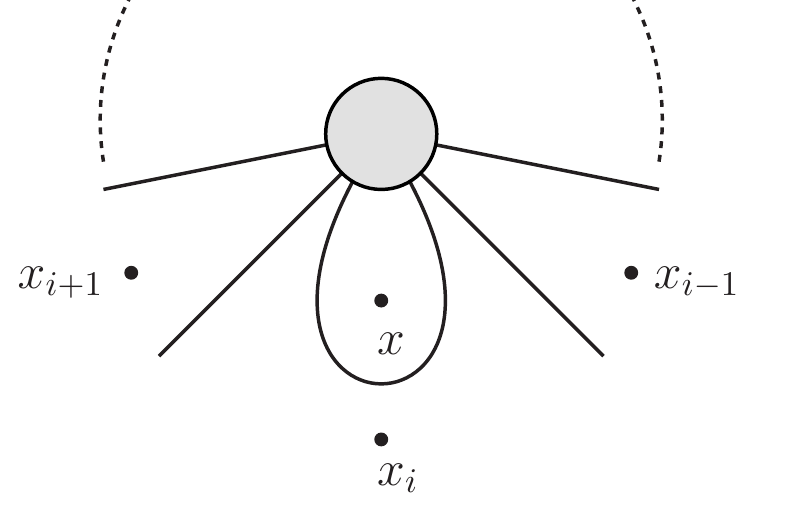}
\caption{\emph{The geometry of the forward limit terms in the recursion relation.}}
\label{fig:ForwardLimit}
\end{figure}

\subsubsection*{Region Momentum Space}

The contribution from to recursion relations from this pole in momentum space is

\begin{equation}
\int \rd^4\eta_I \; \frac{1}{(x-x_i)^2}\; A_{n+2,k+1}^{(\ell-1)}(i,\ldots,i-1,I,-I;z_I)
\end{equation} 
 
\ni where the external region momenta are all shifted and evaluated on the pole $z_I$. Since this is an equation for the loop integrand, the position of the pole $z_I = z_I(x)$ is a function of the internal region momentum. The on-shell momentum $P_I(z_I)$ in the propagator may be written as $\lb_I\tlb_I$ and contracting with the reference spinor we find the holomorphic spinor $\lb_I = P_I\,|\zeta]/[\tlb_I\zeta]$. However, since the forward term is invariant under little group rescaling $(\lb_I,\eta_I)\rightarrow(t \lb_I, t^{-1}\eta_I)$ then we may equally use
\begin{equation}
\label{CSWspinorforward}
\lb_I^{\al} = P_I|\zeta] =  (x-x_i)|\zeta]
\end{equation}

\ni which is again the holomorphic CSW spinor associated with the off-shell propagator. Again, the notation $I$ and $-I$ is shorthand for the variables $\{ \lb_I,\eta_I\}$ and $\{ -\lb_I,\eta_I\}$. Since we are discussing the loop integrand, the spinor $\lb_I=\lb_I(x)$ is really a function of the internal region momentum.

For the loop amplitude itself, the forward terms may be written as dispersion integrals. We may invert the equations $z_I = z_I(x)$ and $\lb_I=\lb_I(x)$ and the express the internal region momentum as a function of the null momentum $\ell^{\al\dal} \equiv \lb_I^{\al}\tlb_I^{\dal}$ and the shift parameter $z$. The integration measure and the propagator then become

\begin{equation}
\frac{\rd^4x}{(x-x_i)^2} = \frac{\rd z}{z} \rD^3\ell
\end{equation}

\ni where $\rD^3\ell = \la \lb_I\rd\lb_I \ra\, \rd^2\tlb_I$ is the standard measure on the null cone. The forward term may now be written as follows

\begin{equation}
\int \frac{\rd z}{z}\rD^3\ell \; \rd^4\eta\;  A_{n+2,k+1}^{(\ell-1)}(i,\ldots,i-1,\{\ell,\eta\},\{-\ell,\eta\}; z)\, .
\end{equation}

\ni which is a dispersion integral in the shift parameter of a single cut of the original amplitude. This formulation of the forward terms reflects their origin from branch cuts of the shifted amplitude in the complex $z$-plane.

\subsubsection*{Momentum Twistor Space}

\begin{figure}[htp]
\centering
\includegraphics[width=14cm]{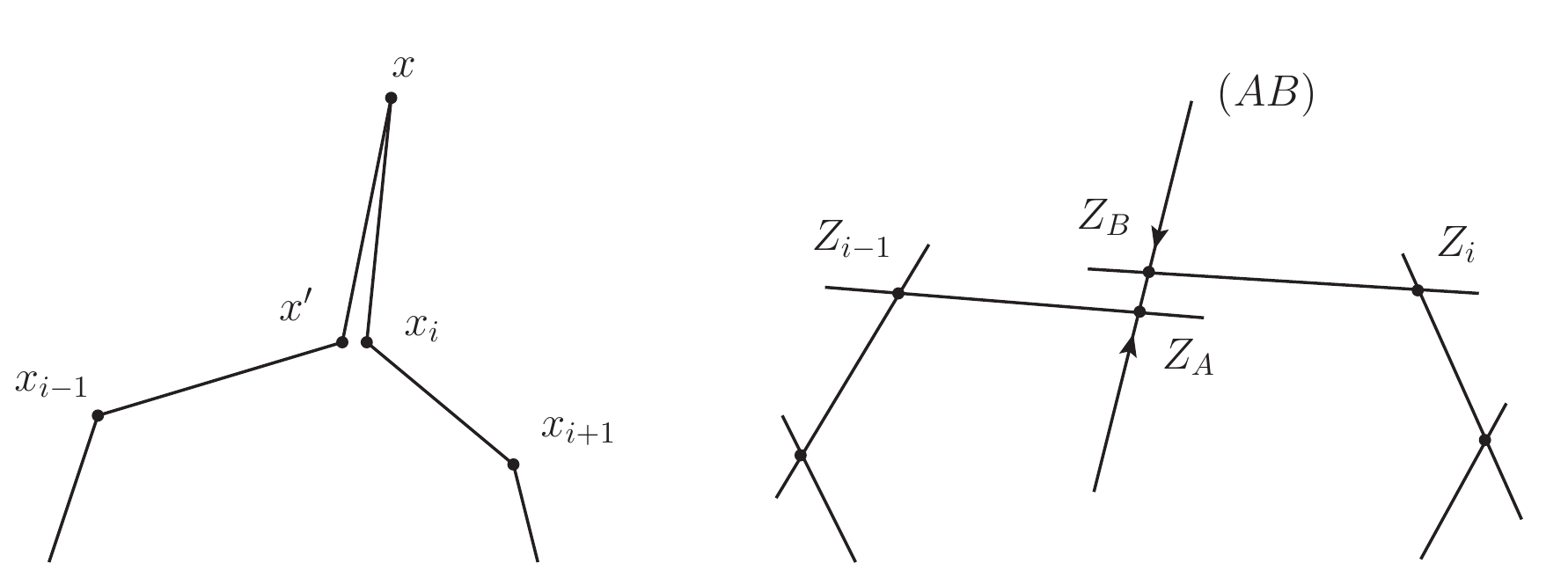}
\caption{\emph{The region space and momentum twistor geometry of a standard forward limit.}}
\label{fig:Forwardgeometry1}
\end{figure}

Let us now understand the momentum twistor geometry of the forward terms. First consider the forward amplitude obtained by cutting the propagator $(x-x_i)^2$ without any shifts of external particles. Starting with an amplitude with $(n+2)$ particles and the additional external regions $(x',x)$ then consider the forward limit where $x'$ tends towards the external region $x_i$ and $x$ becomes a new internal region. In momentum twistor space, then the line $(AB)$ intersects the line $X_i$ in the forward limit and the momentum twistors $Z_A$ and $Z_B$ tend towards the intersection point $(AB)\cap X_i$ (see figure~\ref{fig:Forwardgeometry1}).

Now consider the same forward amplitude with all of the momentum twistors shifted as in equation~\eqref{Momtwistorshift}. Then the shifted line $X_i(z)$ remains in the plane $(\im\,*)$. The pole in the integrand occurs when this line meets the intersection point $B' \equiv (AB)\cap(\im\,*)$ which has components
\begin{equation}
Z_{B'} = \la\,*\,\im\,[Z_A\ra\,Z_B]\, .
\end{equation}

\ni The primary component of $B'$ is then the holomorphic CSW spinor $\lb_I$ associated with the propagator. The forward limit again corresponds to moving the momentum twistors $Z_A$ and $Z_B$ along the line $(AB)$ towards the intersection point $B'$ as illustrated below in figure~\ref{fig:Forwardgeometry2}.

\begin{figure}[htp]
\centering
\includegraphics[width=8cm]{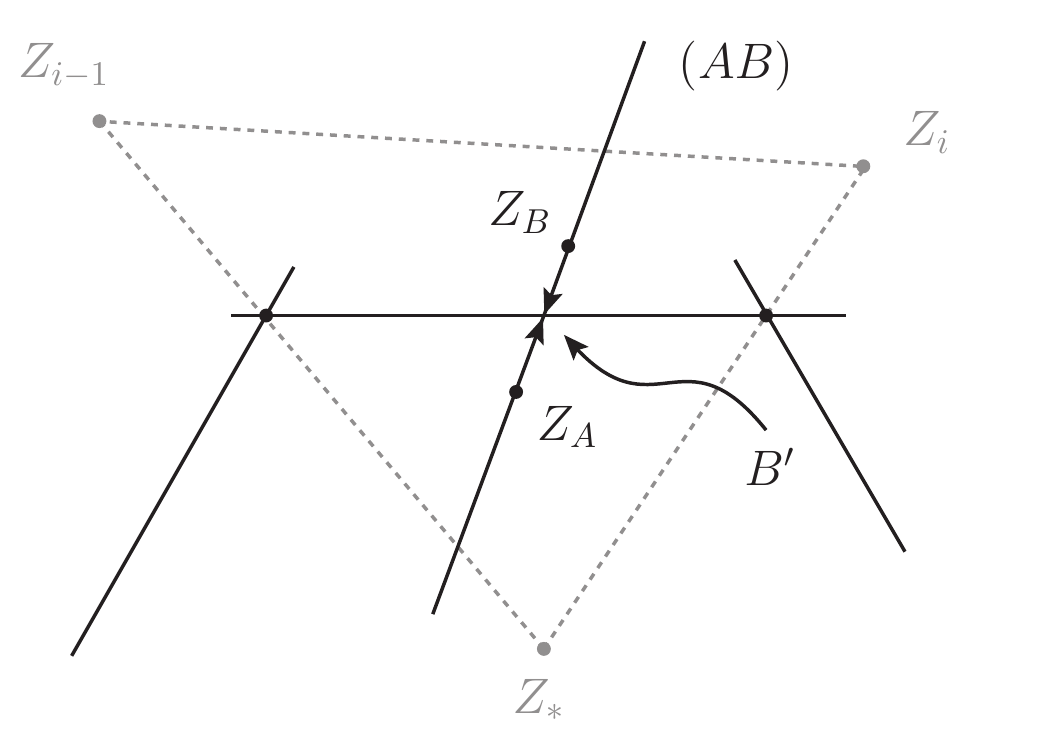}
\caption{\emph{The momentum twistor geometry of forward terms in the all-line recursion relation.}}
\label{fig:Forwardgeometry2}
\end{figure}

The forward limit may be implemented in momentum twistor space following the techniques introduced in~\cite{ArkaniHamed:2010kv}. The dual superconformal loop integration may be decomposed into an integration over lines $(AB)$ and over $\GL(2)$ transformations that move $Z_A$ and $Z_B$ along the line $(AB)$. This can be made explicit by choosing the contour

\begin{equation}
Z_A = (\lb_{A\al},x^{\al\dal}\lb_{A\al}) \qquad Z_B = (\lb_{B\al},x^{\al\dal}\lb_{B\al})
\end{equation}

\ni upon which the dual conformal measure decomposes

\begin{eqnarray}
\rD^{3}Z_A\, \rD^{3}Z_B &=& \rd^4 x \; \la \lb_A\lb_B \ra^2 \la \lb_A \rd \lb_A \ra\la \lb_B\rd \lb_B \ra  \nn\\
&\equiv& \left[ \frac{\rd^4 Z_A \rd^4Z_B}{\mathrm{vol}\,\GL(2)} \right] \; d\mu_{\GL(2)}\, .
\end{eqnarray}

\ni When the integrand depends only on the line $(AB)$ then the contour for the $\GL(2)$ integral may be chosen as the anti-diagonal contour $\lb_A = \overline{\lb}_B$ in $\CP^1\times\CP^1$ and integrates to unity. However, we may also consider integrands with simple poles in $\CP^1\times\CP^1$ arising from spurious denominator factors and in this case we must choose a different contour for the $\GL(2)$ integral.

The additional $\GL(2)$ contour integral may now be employed to perform the forward limit algebraically in momentum twistor space. It is clear from the geometry that the $\GL(2)$ contour must be chosen so $Z_A$ and $Z_B$ are sent to the intersection point $(AB)\cap(\im\,*)$ in order to perform the forward limit.

The contribution to the recursion from the forward term is

\begin{equation}
 \oint\limits_{\mathrm{GL}(2)} \rd\mu \; [A,B,i\!-\!1,i,\,*\,]\; A_{n+2,k+1}^{(\ell-1)}(i\ldots,i-1,A,B';z_I)
\end{equation}

\ni where the external momentum twistors are evaluated on the position of the pole $z_I$. Note that the propagator again contributes a fundamental invariant $[A,B,i\!-\!1,i,\,*\,]$.

\subsubsection*{Vanishing Forward Terms}

It will be important for obtaining MHV diagrams from the recursion relation that the forward terms involving two particles on the same MHV vertex vanish. In each forward term, there is a sum over helicities across the channel, which is implemented by the fermionic integration $\int \rd^4\eta_I$. While the separate component amplitudes diverge as particles become collinear $\lb_1 \propto \lb_2$, the forward limit vanishes as $\cO(\la\lb_1\lb_2\ra)$ once the fermionic integration has been performed. Such a cancellation happens in all supersymmetric theories.

\begin{figure}[htp]
\centering
\includegraphics[width=6cm]{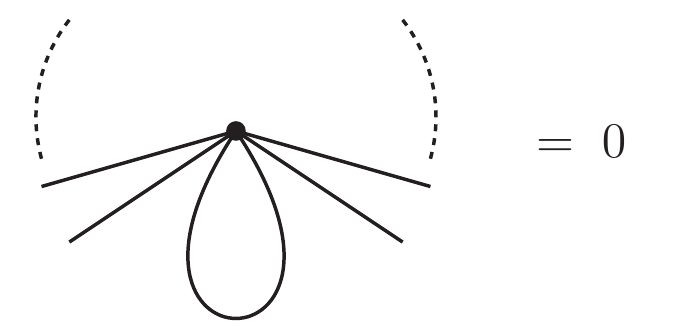}
\caption{\emph{The forward limit of an MHV vertex vanishes in supersymmetric gauge theories.}}
\label{fig:ForwardLimitMHVzero}
\end{figure}


\section{Tree-Level Amplitudes}
\label{sec:TreeLevel}

The solution to the all-line recursion relation~\eqref{RecursionRegion} has been shown to generate the MHV diagram expansion for all tree amplitudes in $\cN=4$ SYM~\cite{Elvang:2008vz}. In the final section, we will prove the corresponding result for all tree and loop amplitudes. Here we take the opportunity to present some tree-level examples of the momentum twistor recursion relation:

\begin{equation}
\label{RecursionTwistorTree}
A_{n,k}^{(0)}(1,\ldots,n) = \sum\limits_{i,j,k_1} [\,*\,,i\!-\!1,i,j\!-\!1,j]\; A_{n_1,k_1}^{(0)}(Z_I,i,\ldots,j-1;z_I )\; A_{n_2,k_2}^{(0)}(Z_I,j,\ldots,i-1;z_I)
\end{equation}

\ni The summation ranges are $0\leq i<j\leq n$ and $0\leq k_1<k-1$ with $k_1+k_2=k+1$ and $n_1+n_2 = n+2$.


\subsection{NMHV}

For tree-level NMHV superamplitudes there is no further work to be done. The only terms in the recursion relation have two MHV vertices connected by a single propagator. Summing over all factorisation channels we have

\begin{equation}
\label{NMHVrecursion}
A_{n,1}^{(0)}(1,\ldots,n) = \sum\limits_{1\leq i<j \leq n}\,[\,*\, , i\!-\!1, i, j\!-\!1, j]
\end{equation}

\ni which is the form of the MHV vertex expansion in momentum twistor space~\cite{Bullimore:2010pj}. 

\begin{figure}[htp]
\centering
\includegraphics[width=11.5cm]{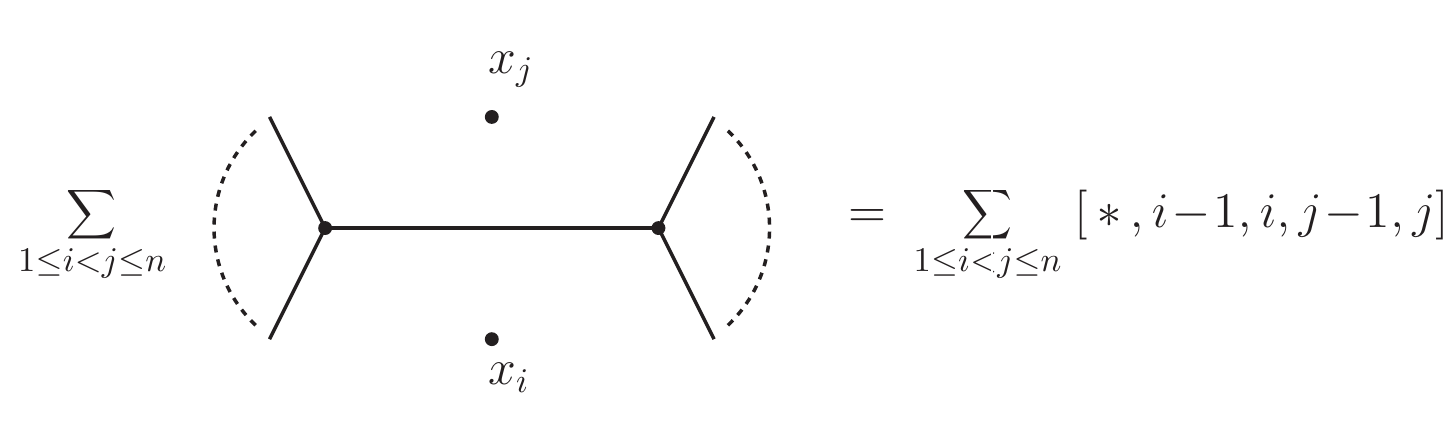}
\caption{\emph{MHV diagrams contributing to the NMHV tree amplitude have a single propagator.}}
\label{fig:NMHVrecursion}
\end{figure}

The dual superconformal symmetry of tree amplitudes~\cite{Drummond:2008vq} means that this expression should be independent of the reference twistor $Z_*$. This can be shown explicitly by expanding each term using linear relations and cancelling terms in pairs~\cite{Bullimore:2010pj} and may also be understood as a grassmannian residue theorem~\cite{ArkaniHamed:2009sx}. When the reference twistor is an external momentum twistor then~\eqref{NMHVrecursion} becomes the BCFW expansion of the amplitude.


\subsection{N$^2$MHV}

Terms in the recursion relation have an MHV and an NMHV vertex connected by a single propagator and the momentum twistor recursion relation becomes

\begin{equation}
\label{NNMHVtreerecursion}
A_{n,2}^{(0)}(1,\ldots,n) = \sum\limits_{1\leq i<j\leq n}\; [\,*\,,i\!-\!1,i,j\!-\!1,j] \; A_{n_2,1}^{(0)}(I,j,\ldots,i\!-\!1;z_I)
\end{equation}

\begin{figure}[htp]
\centering
\includegraphics[width=6.5cm]{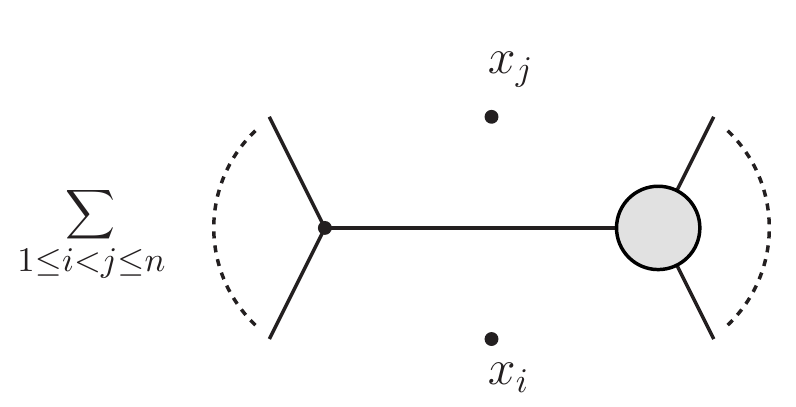}
\caption{\emph{The terms in the recursion relation for the N$^2$MHV tree amplitude. The shaded vertex represents an NMHV tree amplitude.}}
\label{fig:NNMHVrecursion}
\end{figure}

\ni where $n_2 \equiv i\!-\!j$ (see figure~\ref{fig:NNMHVrecursion}). The momentum twistors are all shifted and evaluated on the relevant shift parameter $z_I$ where the momentum $P_I(z) = (x_i-x_j)(z)$ becomes null.

\begin{figure}[htp]
\centering
\includegraphics[width=12.5cm]{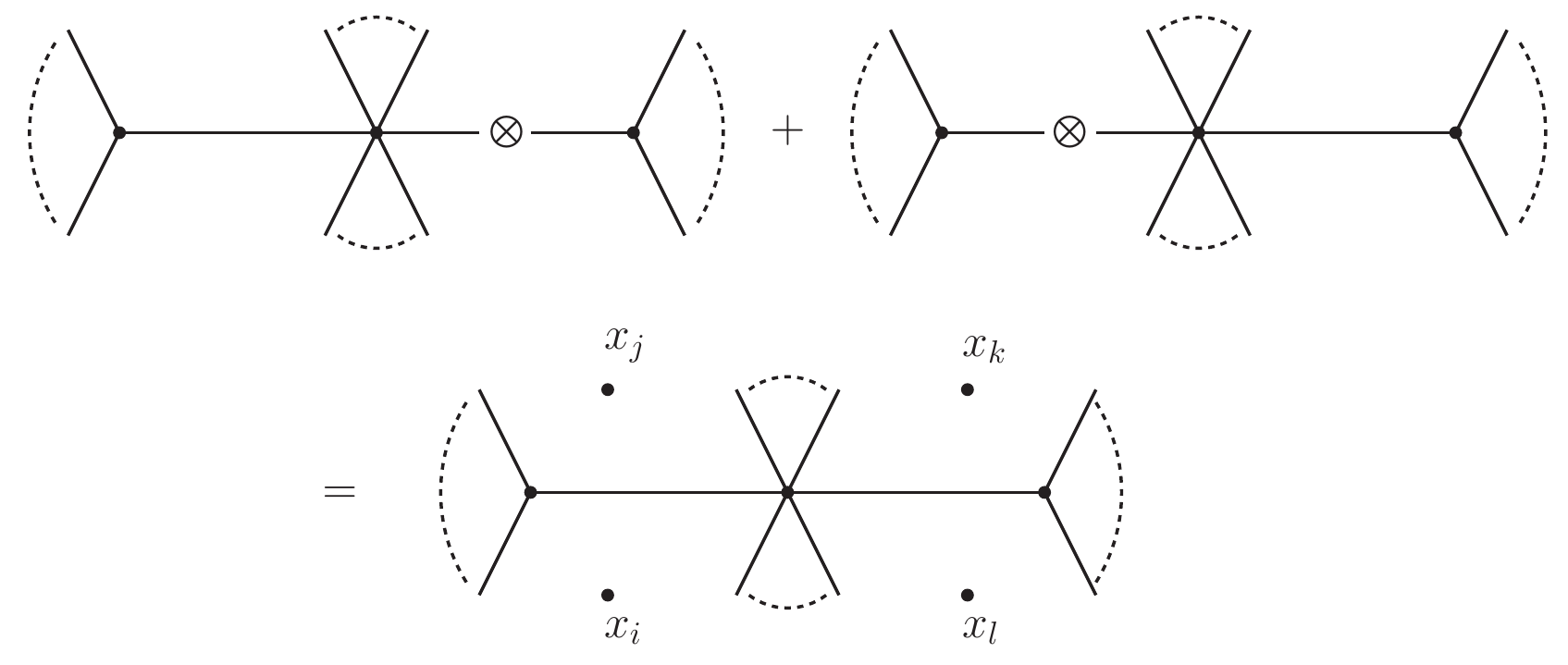}
\caption{\emph{The cancellation of shifted propagators between terms in the recursion.}}
\label{fig:NNMHVtreesum}
\end{figure}

We now expand the NMHV amplitude using the MHV diagram expansion with the same reference twistor $Z_*$. Since all vertices are now MHV, then the momentum twistor shift only effects the propagators in the expansion of the NMHV amplitude, and every term corresponds to an MHV diagram with one propagator shifted and evaluated on the pole of the other. Consider the particular diagram with channels $P_I=(x_i-x_j)$ and $P_J=(x_k-x_l)$ then there are two terms corresponding to this diagram (see figure~\ref{fig:NNMHVtreesum}) whose sum is

\begin{equation}
\label{Nsqsum}
[\,*\, ,i\!-\!1,i,j\!-\!1,j\,]\;[\,*\,,k\!-\!1,k,l\!-\!1,l\,]\; (\frac{P^2_I}{P^2_I(z_J)}+\frac{P^2_J}{P^2_J(z_I)} )  \, . 
\end{equation}

\ni Following~\cite{Elvang:2008vz} we can now imply the contour integral

\begin{eqnarray}
\label{residuetheorem}
0 = \frac{1}{2\pi i} \oint \frac{\rd z}{z} \frac{1}{P_I^2(z) P_J^2(z) } = \frac{1}{P_I^2 P_J^2}  - \frac{1}{P^2_I P^2_J(z_I) } - \frac{1}{P^2_I(z_J) P^2_J}
\end{eqnarray}

\ni to show that the two terms collapse to the correct contribution from the MHV diagram
\begin{equation}
[\,*\, ,i\!-\!1,i,j\!-\!1,j\,]\;[\,*\,,k\!-\!1,k,l\!-\!1,l\,] \, .
\end{equation}

\ni Note that we needed to choose the same reference twistor throughout to generate the MHV diagrams. 
 
Consider now the boundary case where the MHV diagram has channels $P_I=(x_i-x_j)$ and $P_J=(x_k-x_i)$ which are adjacent on the central vertex. There are again two terms in the recursion relation contributing to the MHV diagram, which now involve the equi-anharmonic twistors defined in equation~\eqref{WP1}. For the channels $I$ and $J$ we have
\begin{eqnarray}
Z_I  &=&  Z_{ji} + \frac{1}{2} \la i-1\,i\,j-1\,j \ra Z_* \nn\\
Z_J &=& Z_{ki} + \frac{1}{2}\la k-1\,k\,i-1\,i\ra Z_* \, 
\end{eqnarray}
\ni where the intersections are $Z_{ji} = (j\!-\!1,j)\cap(i\!-\!1,i,\,*\,)$ and $Z_{ki}=(k\!-\!1,k)\cap(i\!-\!1,i,\,*\,)$. Now summing the terms contributing to the given boundary MHV diagram and again using the residue theorem~\eqref{residuetheorem} we find the correct result

\begin{equation}
[\,*\, ,i\!-\!1,i,j\!-\!1,j\,]\;[\,*\,,k\!-\!1,k,i\!-\!1,Z_{ji}]  
\end{equation}

\ni which is the correct expression for the MHV diagram in momentum twistor space.

Summing over all terms in the recursion relation then all MHV diagrams appear twice - once for each channel. Each term is the contribution from an MHV diagram except that one propagator is shifted and evaluated on the pole of the other. The terms then combine in pairs according to the residue theorem~\eqref{residuetheorem} reconstructing each MHV diagram once. The same structure appears in the solution for all tree amplitudes~\cite{Elvang:2008vz} and will appear for the loop amplitudes as well.
\pagebreak


\section{Loop Integrands}
\label{sec:LoopLevel}

In this section, we present iterative solutions of the all-line recursion relation for some one-loop and two-loop amplitudes, which are sufficient to demonstrate the important features. When the same reference spinor is chosen at each stage of the iteration, the solution is the MHV diagram expansion.


\subsection{MHV 1-Loop}

\subsubsection*{A Simple Example}

Before considering the general MHV one-loop integrand, we will understand how the combinatorics of the recursion relation works for the simplest four-point example. For MHV integrands with any numbers of loops, then the only terms in the recursion relation are the forward terms. For the four-point one-loop integrand there are four forward terms arising from propagators $(x-x_1)^2$, $(x-x_2)^2$, $(x-x_3)^2$ and $(x-x_4)^2$ as illustrated below.
\begin{figure}[htp]
\centering
\includegraphics[width=12cm]{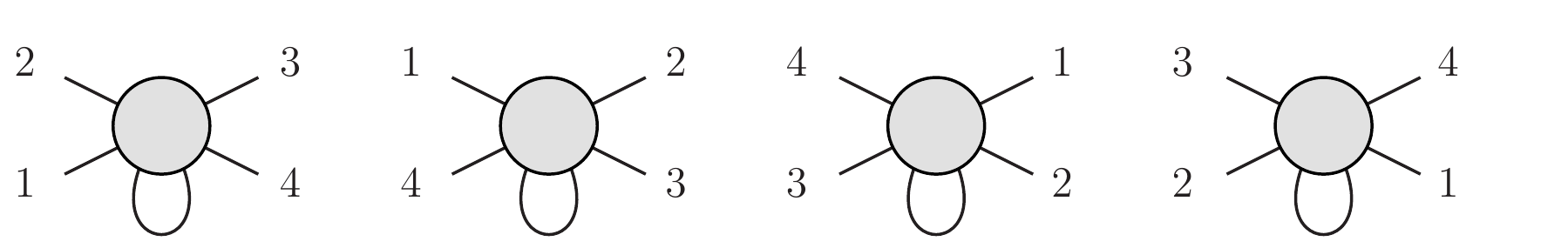}
\label{fig:4pointMHV1}
\end{figure}

The terms involve forward limits of the six-particle NMHV tree amplitude, which may each be expanded in a sum of nine MHV diagrams containing two vertices and a single propagator. However, six of those lead to forward limits in between legs of the same MHV vertex and hence vanish. The remaining three form MHV diagrams of the one-loop MHV integrand except that one propagator has been shifted and evaluated on the pole of the other. For example, for the forward channel $(x-x_1)$
\begin{figure}[htp]
\centering
\includegraphics[width=14cm]{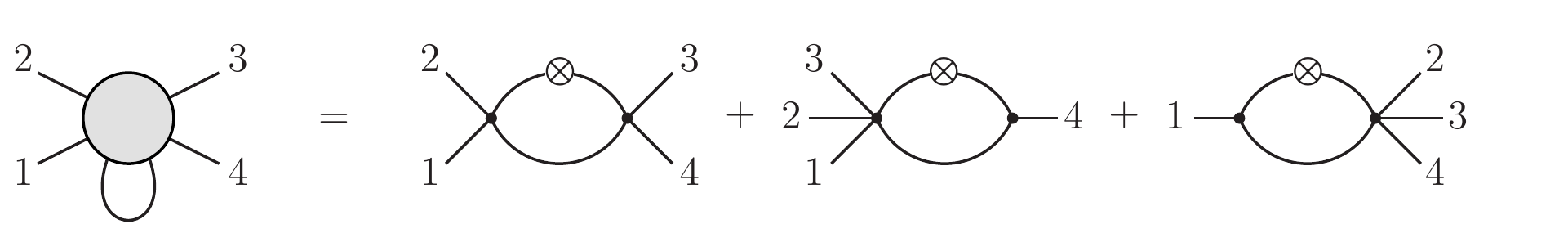}
\label{fig:4pointMHV2}
\end{figure}

There are now twelve terms in the solution of the recursion relation; each of the six diagrams for the one-loop MHV integrands appears twice. However, choosing the same reference spinor at both stages of the recursion, and using simple residue theorems 
\begin{equation}
\oint \frac{ \rd z}{z} \frac{1}{(x-x_i(z))^2(x-x_j(z))^2} = 0
\end{equation}
then the terms combine in pairs to form MHV diagrams. For example, for the MHV diagram with channels $(x-x_1)$ and $(x-x_3)$ we have
\begin{figure}[htp]
\centering
\includegraphics[width=10cm]{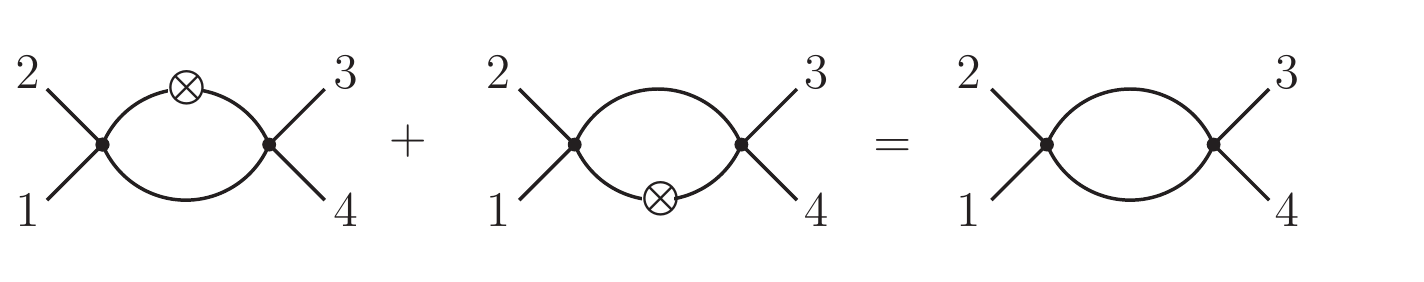}
\label{fig:4pointMHV3}
\end{figure}

\ni In this way all six MHV diagrams are constructed correctly from the solution of the recursion relations. We will now consider the details of this calculation for general one-loop MHV integrands with any number of particles.


\subsubsection*{Region Momentum Space}

\begin{figure}[htp]
\centering
\includegraphics[width=6cm]{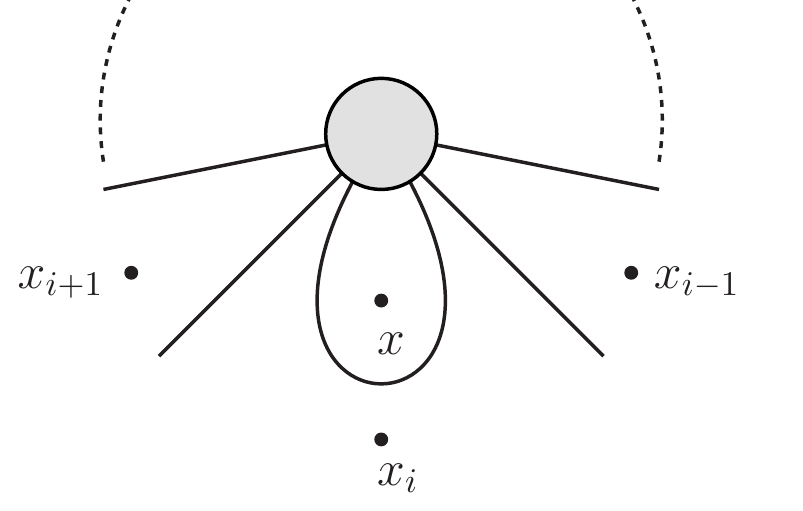}
\caption{\emph{The forward terms contributing to the one-loop MHV integrand. The central vertex is a tree-level NMHV amplitude and $x$ represents the internal region.}}
\label{fig:ForwardMHV}
\end{figure}

Let us first consider the recursion relation for the generic one-loop MHV integrand in region momentum space notation. The only terms in the recursion relation are forward terms and hence the full recursion relation becomes (see figure~\ref{fig:ForwardMHV})

\begin{equation}
A_{n,0}^{(0)}(1,\ldots,n) = \sum\limits_{i=1}^{n} \int \rd^4\eta_I \frac{1}{(x-x_i)^2}A_{n+2,1}^{(0)}(1,\ldots,i-1,\{\lb_I(x),\eta_I\},\{-\lb_I(x),\eta_I\}; z_I(x))
\end{equation}

\ni where $x$ is the single internal region in the one-loop integrand. Since the sub-integrands are not MHV then they depend also on the external region momenta as illustrated in figure~\ref{fig:ForwardMHV}. The poles in the integrand occur when the four-momentum $P_I(z) = (x-x_i(z))$ becomes null
\begin{equation}
z_I(x) =  \frac{(x-x_i)^2}{2\la q_i| (x-x_i) |\zeta]}
\end{equation}

\ni and CSW spinor associated with the propagators $1/P_I^2$ is $\lb_I = P_I|\zeta]$. 
\vspace{0.5cm}

We now expand the tree-level NMHV amplitude with the MHV diagram expansion. The terms involving the forward limit in between two adjacent legs of the same MHV vertex vanish once the fermionic integration has been performed. The remaining terms contain a propagator in an additional channel $(x-x_j)$ and hence the expression for the integrand becomes

\begin{equation}
\label{MHVoneloopregion1}
\sum_{i=1}^n \sum\limits_{j\neq i}\int \rd^4\eta\, \rd^4\eta'\; \frac{A^{(0)}_{n,0}(\{\lb,\eta\},i,\ldots,j-1,\{\lb',\eta'\})A^{(0)}_{n,0}(\{\lb',\eta'\}, j \ldots,i-1,\{  \lb,\eta\} )}{(x-x_i)^2(x-x_j(z_i))^2}
\end{equation}

\ni where the spinors assigned to each propagator are
\begin{equation}
\lb^{\al} = (x-x_i)^{\al\dal}\zeta_{\dal} \qquad \lb'^{\al} = (x-x_j)^{\al\dal}\zeta_{\dal} \, .
\end{equation}

\begin{figure}[htp]
\centering
\includegraphics[width=5cm]{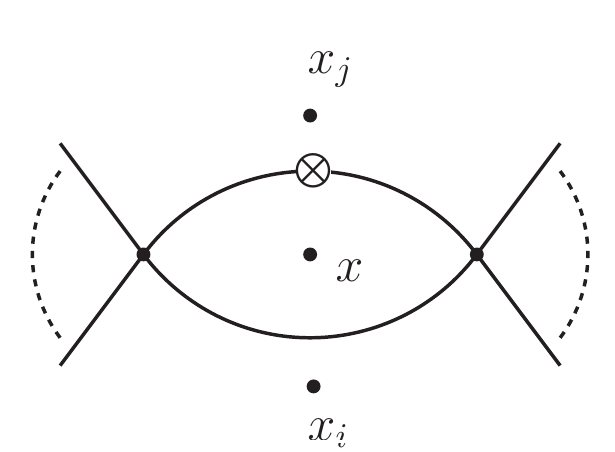}
\caption{\emph{The terms in the recursion summing to an MHV diagram. The tensor product symbol indicates that one propagator is shifted and evaluated on the pole of the other.}}
\label{fig:MHVoneloopsingleterm}
\end{figure}

\ni Since all the external region momenta are shifted and evaluated on $z_i$ then the second propagator from expanding the NMHV amplitude has been shifted to $(x-x_i(z_j))$. However, since the MHV vertices do not depend on the region momenta then this is the only effect of the shift (see figure~\ref{fig:MHVoneloopsingleterm})

\begin{figure}[htp]
\centering
\includegraphics[width=10cm]{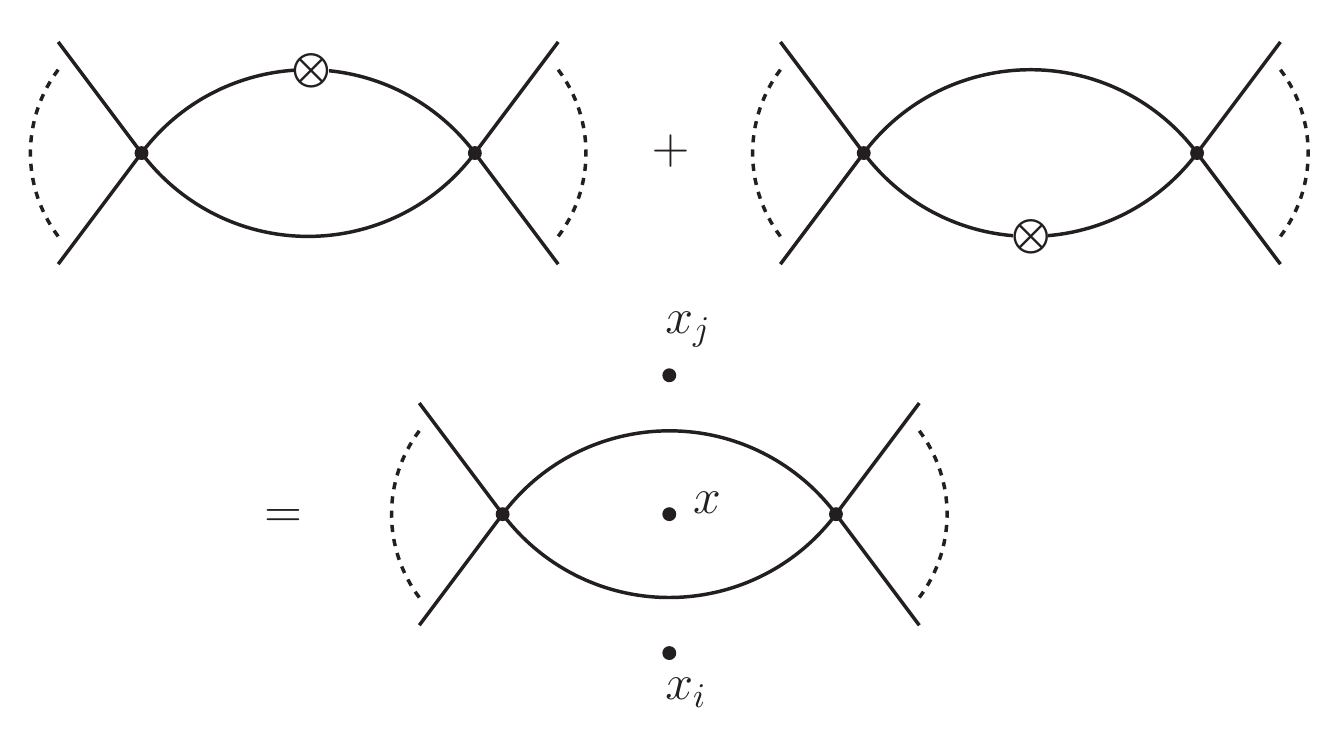}
\caption{\emph{The two terms in the recursion summing to an MHV diagram. Tensor product symbols represent propagators shifted and evaluated on the pole of the other.}}
\label{fig:MHVoneloopterms}
\end{figure}

Each term in the expansion is the exactly the contribution from an MHV diagram except for the shifted propagator. Every MHV diagram for the one-loop integrand then appears in the recursion relation twice - once for each channel (see figure~\ref{fig:MHVoneloopterms}). Consider now the particular MHV diagram with channels $(x-x_i)$ and $(x-x_j)$ then summing the two contributing terms in the recursion relation we find the correct contribution 

\begin{equation}
\int \rd^4\eta\, \rd^4\eta'\; \frac{A^{(0)}_{n,0}(\{\lb,\eta\},i,\ldots,j-1,\{\lb',\eta'\})A^{(0)}_{n,0}(\{\lb',\eta'\}, j \ldots,i-1,\{  \lb,\eta\} )}{(x-x_i)^2(x-x_j)^2}\, .
\end{equation}

\ni multiplied by an overall factor

\begin{equation}
\label{MHVfactor}
\left[ \frac{(x-x_i)^2}{(x-x_i(z_j))^2} + \frac{(x-x_j)^2}{(x-x_j(z_i))^2} \right]\, .
\end{equation}

\ni However, following the discussion of the tree-level amplitudes we consider the following contour integral where the contour surrounds all poles in the integrand

\begin{equation}
\label{MHVoneloopcontour}
\oint \frac{ \rd z}{z} \frac{1}{(x-x_i(z))^2(x-x_j(z))^2} = 0
\end{equation}

\ni which implies that the expression~\eqref{MHVfactor} is simply unity. Summing all of the terms in the recursion relation we find each MHV diagram appears once in the summation:

\begin{equation}
\sum\limits_{1 \leq i < j \leq n} \int \rd^4\eta\, \rd^4\eta'\; \frac{A^{(0)}_{n,0}(\{\lb,\eta\},i,\ldots,j-1,\{\lb',\eta'\})A^{(0)}_{n,0}(\{\lb',\eta'\}, j \ldots,i-1,\{  \lb,\eta\} )}{(x-x_i)^2(x-x_j)^2}\, .
\end{equation}

Consider now the one-loop amplitude itself. We can then express the MHV diagrams as dispersion integrals which reconstruct the standard box expansion from its unitarity cuts~\cite{Brandhuber:2004yw}. We first introduce an additional integration variable 

\begin{equation}
\frac{\rd^4x}{(x-x_i)^2(x-x_j)^2} = \frac{\rd^4x}{(x-x_i)^2}\frac{\rd^4y}{(y-x_j)^2} \delta^4(x-y)
\end{equation}

\ni and decompose each of them using the reference spinor

\begin{equation}
(x_1-x_i) = \ell_i + z_i q \qquad (x_2-x_j) = \ell_j + z_j q
\end{equation}

\ni where the null reference momentum is $q^{\al\dal} = \zeta^\al \tilde{\zeta}^{\dal}$ and the null four-momenta $(\ell_i,\ell_j)$ and the parameters $(z_i,z_j)$ are defined by, for example,

\begin{equation}
|\ell_i\ra = (x_1-x_i)|\iota] \qquad z_i = \frac{2q\cdot(x_1-x_i)}{(x_1-x_i)^2}\, .
\end{equation}

\ni The integration measure may then be rewritten in the new variables

\begin{eqnarray}
\int \frac{\rd^4x_1}{(x_1-x_i)^2}\frac{\rd^4x_2}{(x_2-x_i)^2} \delta^4(x_1-x_2) &=& \int \frac{\rd z_1}{z_1}\frac{\rd z_2}{z_2}\rD^3\ell_1 \rD^3\ell_2\; \delta^4(x_{ij} + \ell_i - \ell_j + (z_i - z_j)q) \nn\\
&=& \int \frac{\rd z}{z} \rd\mathrm{LIPS}(\ell_i,-\ell_j,x_{ij}(z) )
\end{eqnarray}

\ni where we have defined $z=(z_i-z_j)$ and $x_{ij}(z) = x_{ij}-zq$. Therefore expanding the NMHV vertex has turned each dispersion integral of a single-cut into a sum of dispersion integrals over standard unitarity cuts. The dispersion integrals then reconstruct the standard box expansion from its unitarity cuts as shown in~\cite{Brandhuber:2004yw}.
\pagebreak

\subsubsection*{Momentum Twistor Space}

Let us now perform the same calculation using the momentum twistor form of the recursion relation. Again, the only terms involve forward limits on tree-level NMHV superamplitudes. Hence we have the following expression for the momentum twistor integrand
\begin{equation}
\label{MHVonelooprecursion}
A_{n,0}^{(1)}(1,\ldots,n) = \sum\limits_{i=1}^{n}\; \oint\limits_{\GL(2)}\; [\,*\,,i\!-\!1,i,A,B]\; A_{n+2,1}^{(0)}(i,\ldots,i\!-\!1,A,B';z_i)
\end{equation}

\ni where the contour is chosen in order to perform the forward limit in where $Z_A$ and $Z_B$ are sent to the intersection $B'=(AB)\cap(i\!-\!1,i,\,*\,)$. We now expand the tree-level NMHV integrand using the MHV diagram expansion. Terms with forward limits involving two legs on the same vertex are independent of either $\chi_A$ or $\chi_B$ or both, and hence vanish upon fermionic integration. The remaining terms involve additional channels $(x-x_j)$ and the integrand becomes

\begin{equation}
\label{MHVonelooprecursion2}
\sum\limits_{i=1}^{n} \sum\limits_{j\neq i} \frac{(x-x_j)^2}{(x-x_j(z_i))^2}\; \oint\limits_{\GL(2)} \rd\mu\; [\,*\,,i\!-\!1,i,A,B]\; [\,*\,,j\!-\!1,j,A,B'] \, .
\end{equation}

\ni The all-line shift only effects the denominator factor $\la A\,B\,j\!-\!1\,j\ra$ corresponding to the propagator $(x-x_j)^2$ and hence the dependence has been pulled outside the $\GL(2)$ integral. 

We now perform the $\GL(2)$ contour integral which implements the forward limit. The contour encloses the poles in the integrand arising from the denominator factors $\la A\,\im\,*\,\ra$ and $\la B\,\jm\,*\,\ra$ and therefore we then have the following contour integral

\begin{equation}
\oint \frac{\la \lb_A\rd\lb_A \ra\la \lb_B\rd\lb_B \ra}{\la \lb_A\lb_B \ra\la A\,\im\,*\,\ra\la B\,\jm\,*\,\ra} =  \frac{1}{\la\, *\, \im \,[A\,\ra\la\, B]\, \jm\,*\,\ra}\, .
\end{equation}

\ni The result of the contour integral may be absorbed by introducing the intersection point $B'' = (AB)\cap(j-1\,j\,*)$ and replacing the unshifted invariant by $[\,*\,,i\!-\!1,i,A,B'']$. The expression for the loop integrand now becomes

\begin{eqnarray}
\label{MHVonelooprecursion3}
\sum\limits_{i-1}^n \sum\limits_{j\neq i}\frac{(x-x_j)^2}{(x-x_j(z_i))^2}  \; [\,*\,,i\!-\!1,i,A,B'']\; [\,*\,,i\!-\!1,i,A,B']\, .
\end{eqnarray}

\ni Each MHV diagram now appears twice and we use the same residue theorem to combine the terms. The result for the momentum twistor integrand is then

\begin{equation}
A_{n,0}^{(1)}(1,\ldots,n) = \sum\limits_{1\leq i<j\leq n}  [\,*\,,i\!-\!1,i,A,B'']\; [\,*\,,i\!-\!1,i,A,B']\, .
\end{equation}

\ni It has been checked numerically that this result is independent of $Z_*$ and shown explicitly that it reproduces the standard box expansion of the amplitude~\cite{Bullimore:2010pj}. Choosing the $Z_*=Z_1$  immediately reproduces the BCFW expression for the loop integrand~\cite{ArkaniHamed:2010kv}.


\subsection{NMHV 1-Loop}


\subsubsection*{A Simple Example}

Firstly, we consider schematically the simplest example with five particles. There are now both factorisation and forward terms in the recursion relation. The factorisation terms lead to diagrams with a one-loop MHV sub-diagram connected to an MHV vertex. For example, the factorisation term in the channel $(x_1-x_3)$ leads to
\begin{figure}[htp]
\centering
\includegraphics[width=15cm]{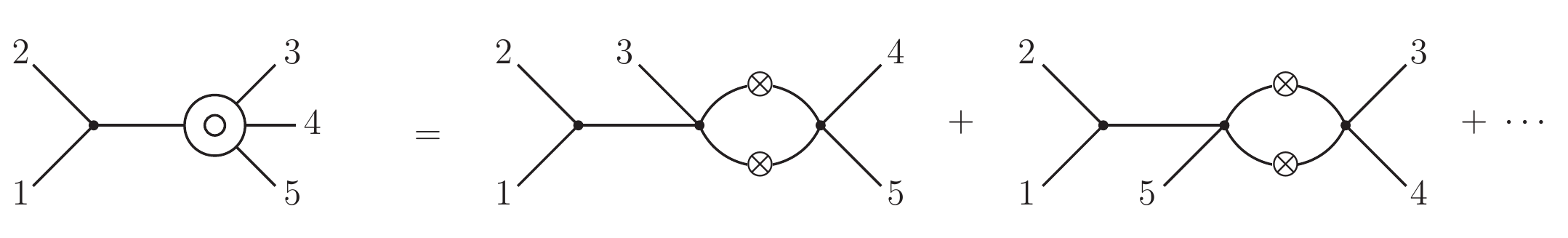}
\end{figure}

\ni In each diagram, two propagators are shifted and evaluated on the pole of the third. The forward terms involve forward limits of the $\Nsq$ tree amplitude and lead also to triangle diagrams. For example, for the forward term with channel $(x-x_1)$ 
\begin{figure}[htp]
\centering
\includegraphics[width=11.5cm]{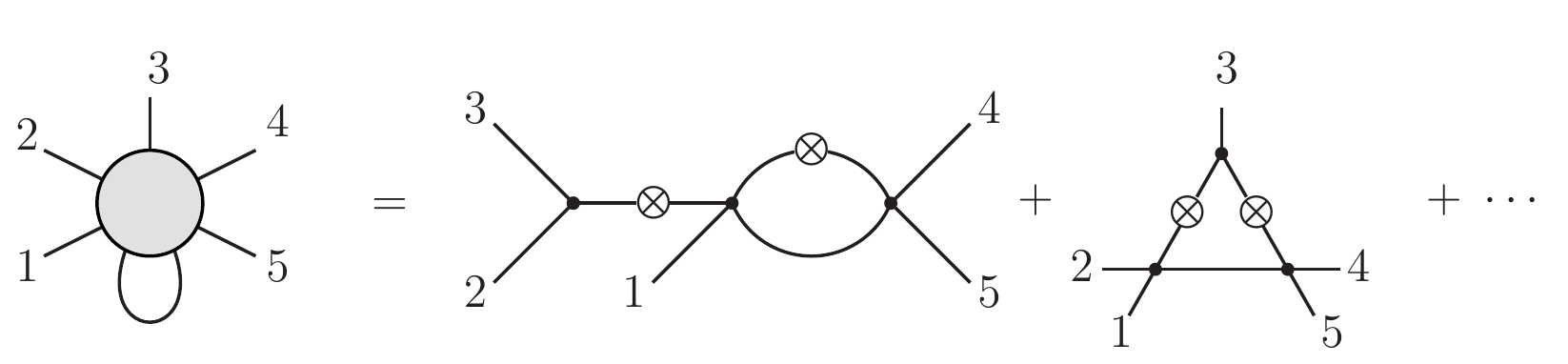}
\end{figure}

Summing over all factorisation and forward terms then each MHV diagram for the one-loop NMHV integrand appears three times - once for each propagator. Residue theorems then ensure that the terms combine in threes to produce each MHV diagram once.


\subsubsection*{General 1-Loop NMHV Integrands}

\begin{figure}[htp]
\centering
\includegraphics[width=12cm]{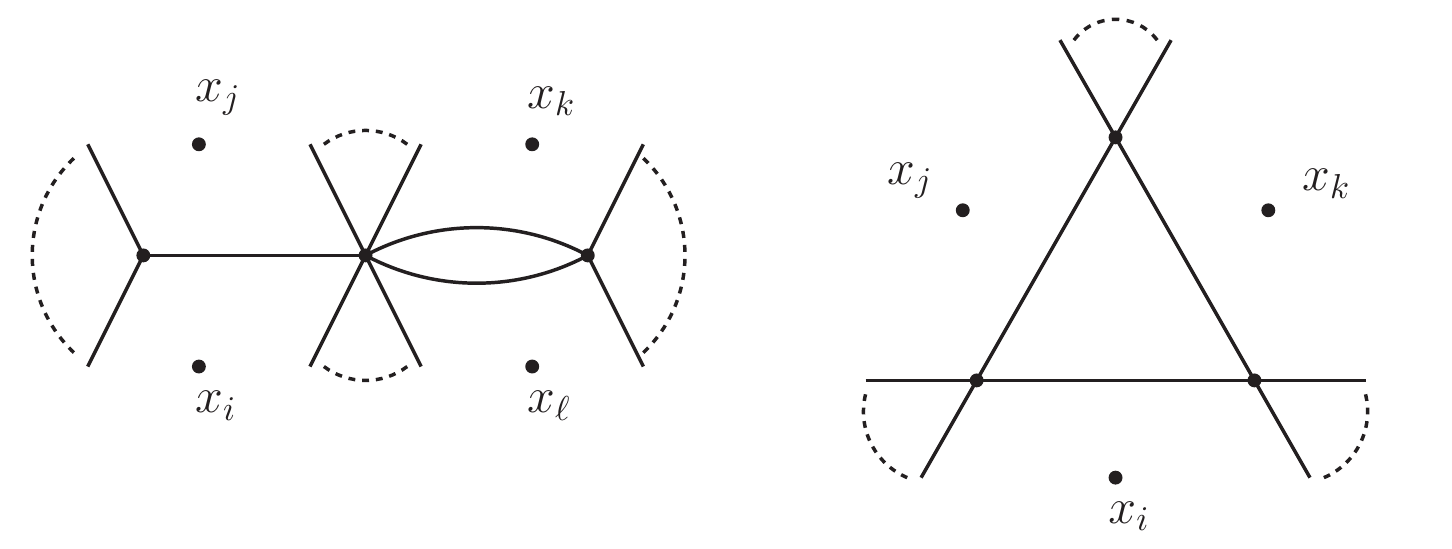}
\caption{\emph{The two classes of MHV diagram for the one-loop NMHV integrand.}}
\label{fig:TheNMHVoneloopMHVdiagrams}
\end{figure}

For 1-loop NMHV integrands, there are both forward terms and factorisation terms in the recursion relation, and two classes of MHV diagrams that should arise from the recursion relation (see figure~\ref{fig:TheNMHVoneloopMHVdiagrams}). 

The recursion relation for the momentum space integrand is

\begin{eqnarray}
A_{n,1}^{(1)}(1,\ldots,n) &=& \sum\limits_{1\leq i<j \leq n}\; \int \rd^4\eta_I \; A_{n_1,0}^{(0)}(i,\ldots,j,I;z_I)\; \frac{1}{(x_i-x_j)^2}\; A_{n_2,0}^{(1)}(j,\ldots,i-1,I;z_I) \nn\\
&+& \sum\limits_{i=1}^n \frac{1}{(x-x_i)^2}\int \rd^4\eta_I \; A_{n+2,2}^{(0)}(i,\ldots,i-1,I,-I;z_I)\, .
\end{eqnarray}

\ni and similarly for the momentum twistor integrand 

\begin{eqnarray}
A_{n,1}^{(1)}(1,\ldots,n) &=& \sum\limits_{1\leq i<j\leq n} [\,*\,,i\!-\!1,i,j\!-\!1,j]\; A_{n_2,0}^{(1)}(Z_I,j,\ldots,i-1;z_I) \nn\\
&+&\sum\limits_{i=1}^{n}\; \oint \rd\mu_{\mathrm{GL}(2)}\; [\,*\,,i\!-\!1,i,A,B] \; A_{n+2,2}^{(0)}(i,\ldots,i-1,A,B';z_i) \, . 
\end{eqnarray}

\begin{figure}[htp]
\centering
\includegraphics[width=13cm]{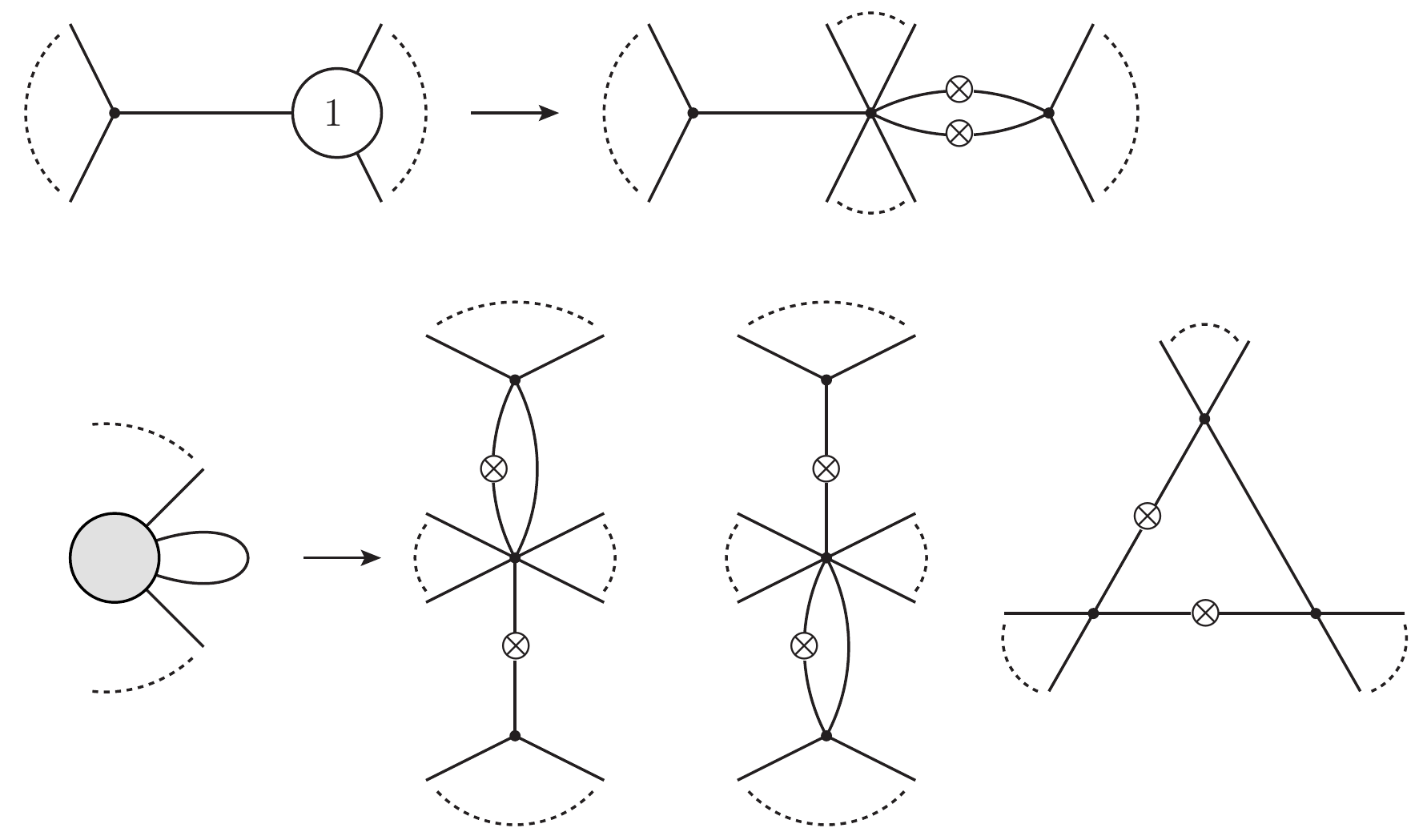}
\caption{\emph{The two terms in the recursion summing to an MHV diagram.}}
\label{fig:NMHVoneloopterms}
\end{figure}

\ni The generic terms found by expanding the integrands are illustrated in figure~\ref{fig:NMHVoneloopterms}. Each term is an MHV diagram where two propagators are shifted and evaluated on the pole of the third. Hence each MHV diagram appears three times in the expansion - once for each propagator. Residue theorems of the form

\begin{equation}
\oint \frac{\rd z}{z} \frac{1}{P^2_I(z)P^2_J(z)P^2_K(z)}=0
\end{equation}

\ni where each propagator $1/P_I^2$ may be either bound two external regions $1/(x_i-x_j)^2$ or bound an external and an internal region $1/(x-x_i)^2$. Then the solution of the recursion relation generates every MHV diagram once.

\subsubsection*{Triangular Diagrams}

Here we consider another example of computing with the momentum twistor recursion relation, which differs from the one-loop MHV case considered above. The triangular MHV diagrams all arise from forward terms in the recursion relation, and hence, consider the forward term in the channel $(x-x_i)$ 

\begin{equation}
\label{NMHVonelooprecursion}
\oint \rd\mu_{\GL(2)}\; [A,B,i-1,i,\,*\,]\; A^{(0)}_{n,2}(i,\ldots,i-1,A,B';z_I)
\end{equation}

\ni We now expand the tree-level N$^2$MHV amplitude with the MHV diagram expansion. The triangular diagrams come from boundary terms in the expansion of the N$^2$MHV amplitude. The boundary terms contributing to the integrand $A^{(0)}_{n,2}(i,\ldots,i\!-\!1,A,B')$ are 

\begin{equation}
\sum\limits_{i<j<k<i\!-\!1} [\,*\,,j-1,j,A,B']\; [\,*\,,k-1,k,A,B''].
\end{equation}

\ni where we have the second intersection point 
\begin{equation}
B''=(AB')\cap(j,j\!-\!1,j,\,*\,) = (AB)\cap(j\!-\!i,j,\,*\,)\, .
\end{equation}

Now consider the contributions to the triangular diagram in figure~\ref{fig:TheNMHVoneloopMHVdiagrams}. Since the all-line shift only effects the propagators, then summing the three contributing terms, we have
\begin{equation}
\oint\limits_{\GL(2)}\; [\,*\,,i-1,i,A,B]\;  [\,*\,,j-1,j,A,B']\; [\,*\,,k-1,k,A,B''] 
\end{equation}
multiplied by an overall factor
\begin{equation}
\frac{(x-x_j)^2(x-x_k)^2}{(x-x_j(z_i))^2(x-x_k(z_i))^2}
\end{equation}
\ni The $\GL(2)$ contour integral is taken to surround the poles $\la A\,\im\,*\ra$ and $\la B\,\km\,*\ra$ which performs the forward limit. The outcome of the forward limit is to replace $ [\,*\,,i\!-\!1,i,A,B]$ with the invariant $ [\,*\,,i\!-\!1,i,A,B''']$ where we have defined $B'''=(AB)\cap(i\!-\!1,i,\,*\,)$. The remaining two contributions to the triangle diagram are obtained by cyclically permuting $\{i,j,k\}$. However the following contour integral
\begin{equation}
\oint \frac{\rd z}{z} \frac{1}{(x-x_i(z))^2(x-x_j(z))^2(x-x_k(z))^2} = 0
\end{equation}

\ni means that the sum of three terms combine to become
\begin{equation}
[\,*\,,i\!-\!1,i,A,B''']\;  [\,*\,,j\!-\!1,A,B']\; [\,*\,,k\!-\!1,k,A,B'']
\end{equation}

\ni which is the correct contribution from the triangular MHV diagram in momentum twistor space.





\subsection{MHV 2-Loop}

\subsubsection*{A Simple Example}

Consider now the four-point two-loop MHV integrand. There are four forward terms in the recursion relation - one for each of the channels $(x-x_1),\ldots,(x-x_4)^2$ - each of which involves the forward limit of an NMHV one-loop integrand.
\begin{figure}[H]
\centering
\includegraphics[width=12cm]{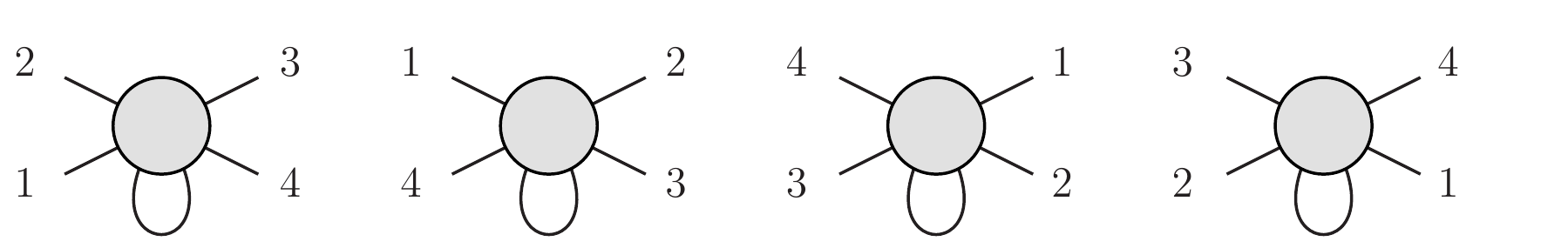}
\end{figure}
However, since there are now two internal regions, we must introduce two region momenta $(y_1,y_2)$ and symmetrise over their assignment to internal regions. Therefore, there are really eight terms and an overall factor of one-half. Expanding the NMHV one-loop integrands then each forward term leads to diagrams of two classes - triangles with bubble sides and double bubbles. For example, in the channel $(x-x_1)^2$ we have an expansion
\begin{figure}[htp]
\centering
\includegraphics[width=12cm]{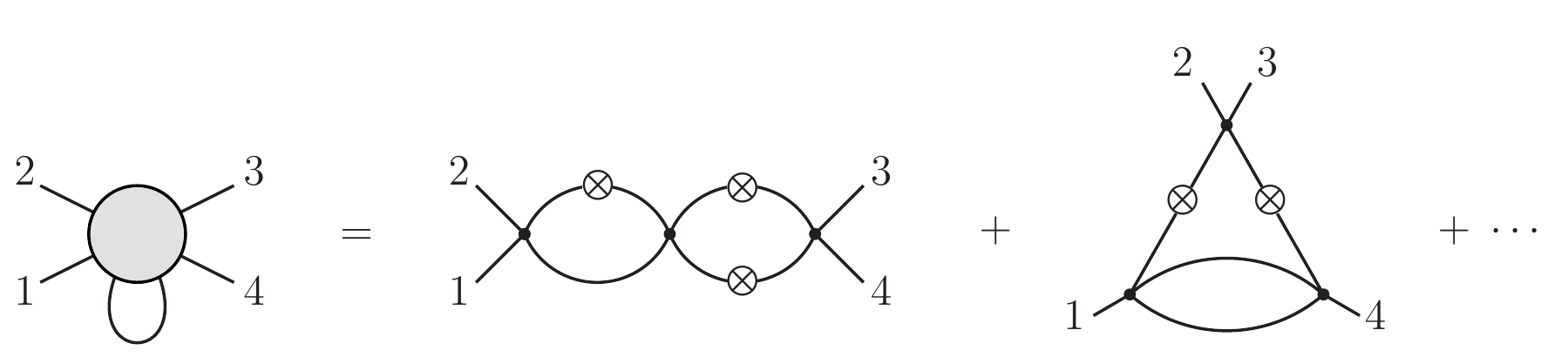}
\end{figure}

We now encounter a new feature starting at two loops. The double bubble diagrams have four channels bounding external and internal regions and hence each one appears four times in the expansion. For the double bubble illustrated above we have the residue theorem
\begin{equation}
\label{residuea}
\int \frac{\rd z}{z} \frac{1}{(y_1-x_1(z))^2(y_2-x_1(z))^2(y_1-x_3(z))^2(y_2-x_3(z))^2} =0
\end{equation}
which combines four terms into the correct MHV diagram. However, the triangle diagrams have an internal channel whose propagator is unshifted, and hence each appears three times in the expansion. For the triangle diagram illustrated above we have the residue theorem

\begin{equation}
\label{residueb}
\int \frac{\rd z}{z} \frac{1}{(y_1-x_1(z))^2(y_2-x_2(z))^2(y_2-x_4(z))^2} =0
\end{equation}
which combines three terms into the correct MHV diagram. The number of times a generic MHV diagram appears in the recursion relation is always equal to the number of channels bounding at least one external region.

\pagebreak

\subsubsection*{General 2-Loop MHV Integrands}

\begin{figure}[htp]
\centering
\includegraphics[width=11cm]{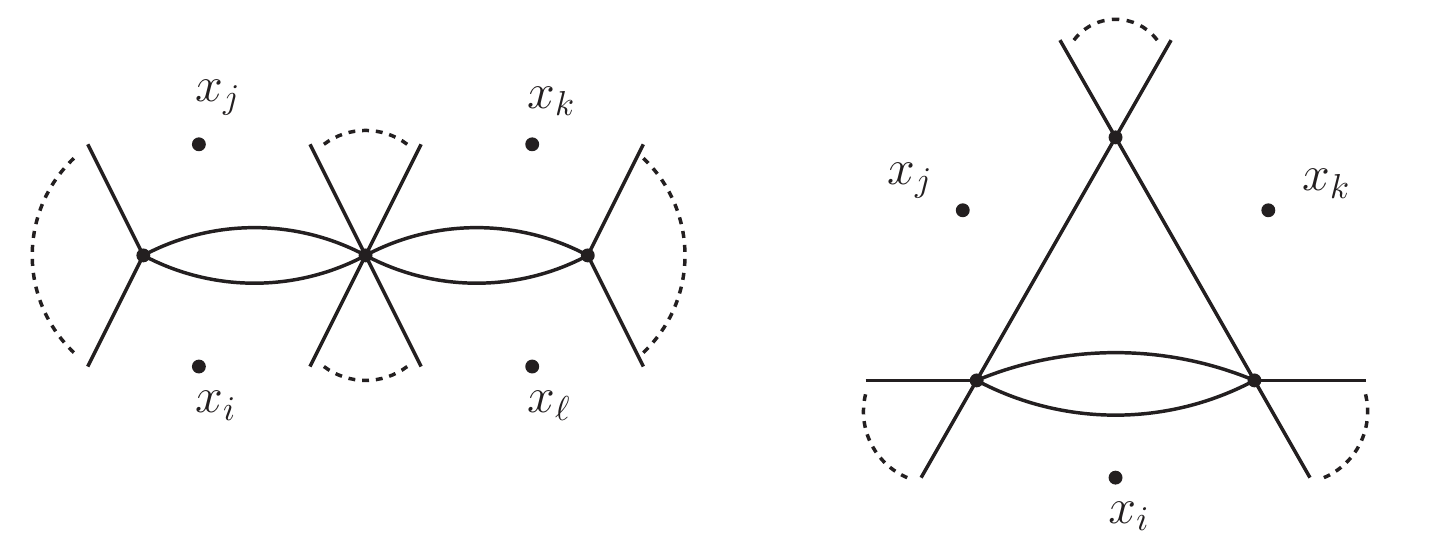}
\caption{\emph{The two classes of MHV diagram for the two-loop MHV integrand.}}
\label{fig:TheMHVtwoloopMHVdiagrams}
\end{figure}

The generic diagrams contributing to an MHV two-loop integrand are shown in figure~\ref{fig:TheMHVtwoloopMHVdiagrams}. In discussing the loop integrand we symmetrise over the lines $(AB)_1$ and $(AB)_2$ in momentum twistor space (or over the internal region momenta $y_1$ and $y_2$). Then diagrams related by $(AB)_1 \leftrightarrow (AB)_2$ are distinct and the summation comes with an overall factor of one-half.

The recursion relation contains only forward terms involving forward limits of one-loop NMHV integrands. In momentum space we have the recursion relation

\begin{equation}
A_{n,0}^{(2)}(1,\ldots,n) =  \sum\limits_{i=1}^n \frac{1}{(y_1-x_i)^2}\int \rd^4\eta\;  A_{n+2,1}^{(1)}(i,\ldots,i-1,I,-I;\,z_i)
\end{equation}

\ni where the one-loop NMHV integrand has an internal region $y_2$ and the whole expression is then to be symmetrised in $y_1$ and $y_2$. In momentum twistor space we have

\begin{equation}
A_{n,0}^{(1)}(1,\ldots,n) = \sum\limits_{i=1}^{n}\; \oint\limits_{\GL(2)} [\,*\,,i-1,i,A,B]\; A_{n+2,1}^{(1)}(i,\ldots,i-1,A,B';z_i)
\end{equation}

\ni where one-loop NMHV integrand depends also on an internal region $(CD)$ and the expression is to be symmetrised in $(AB)$ and $(CD)$. Expanding the NMHV one-loop integrand in MHV diagrams and summing over forward channels $i=1,\ldots,n$ then the double bubble diagrams occur four times and the triangle diagrams occur three times (recall that diagrams related by $(AB)\leftrightarrow(CD)$ are considered separate). Residue theorems of the form~\eqref{residuea} and~\eqref{residueb} then combine terms into MHV diagrams.











\section{Proof of MHV Diagrams for all Loop Integrands}
\label{sec:Proof}

In this section we prove that the solution of the all-line recursion relation (choosing the same reference spinor throughout the recursive solution) is exactly the MHV diagram expansion for all loop integrands in $\cN=4$ SYM. The proof is by induction and extends the argument presented in~\cite{Elvang:2008vz} and simplified recently in~\cite{Cohen:2010mi} for tree-level amplitudes. We note that an MHV diagram for the $\ell$-loop $\Nk$ integrand contains $P\equiv k+2\ell$ propagators.

\subsection*{Outline of Proof}

We will first present an outline of the proof. 

\begin{enumerate}
\item The induction starts with the tree-level NMHV amplitude. We have already shown that the all-line recursion directly generates the MHV expansion of this amplitude.
\item To begin the inductive step, we assume that the MHV expansion correctly reproduces the integrand for all $m$-loop N$^q$MHV amplitudes with 
	\begin{itemize}
	\item $q<k$ and $m\leq\ell$	
	\item $q= k+1$ with $m=\ell-1$\, .
\end{itemize}
\item To complete the inductive step, we consider the all-line recursion relation for the integrand of the $\ell$-loop $\Nk$ amplitude. The recursion relation involves two kinds of terms:
\begin{itemize}
\item Factorisation terms involving $m$-loop N$^q$MHV integrands with $m\leq\ell$ and $q<k$.
\item Forward terms involving the $(\ell-1)$-loop N$^{k+1}$MHV integrand
\end{itemize}
According to our assumption these may be expanded in MHV diagrams. We then show that the recursion relation reproduces the MHV diagram expansion of the $(\ell-1)$-loop N$^{k+1}$MHV integrand.
\end{enumerate}

\ni We will now complete step 3 looking at the factorisation and forward terms in turn.


\subsection*{Factorisation Terms}

So consider the all-line recursion relation for the $\ell$-loop N$^k$MHV superamplitude. We first consider the standard factorisation terms from the channel $I$ with momentum $P_I = (x_i-x_j)$ bounding two external regions. Consider the term where the integrands on either side of the propagator are N$^q$MHV and N$^{k-q+1}$MHV and have $m$ and $(\ell-m)$ loops respectively, then the contribution to the recursion relation is

\begin{equation}
\label{RecursionFactorisation}
\int \rd^4\eta_I\; A_{q}^{(m)}(i,\ldots,j-1,I;z_I) \; \frac{1}{P_I^2} \; A^{(\ell-m)}_{k-q-1}(j,\ldots,i-1,-I,z_I)\, .
\end{equation}
 
\ni These terms must be summed over the degrees of the integrands $(0 \leq q \leq k-1)$, the numbers of loops $(0 \leq m \leq \ell)$ and symmetrised over the assignment of all internal region momenta. All standard factorisation channels are then obtained by summing over the range $(1\leq i < j \leq n)$. The terms where $i$ and $j$ are separated by fewer than two (modulo $n$) will vanish automatically.

We now replace the sub-integrands in the factorisation terms with their MHV diagram expansions using the same reference spinor $\zeta^{\dal}$ as in the recursion relation. Each term in the expansions of the sub-integrands depends on the shifted internal momentum $P_I(z_I) = \lb_I \tlb_I$ only through the holomorphic CSW spinor $\lb_I  = P_I |\zeta]$. Hence every term in the expansion corresponds to an MHV diagram for the original $\ell$-loop $\Nk$ integrand. However, from the all-line shift of external region momenta, then all propagators bounding external regions (except $1/P_I^2$) are shifted and evaluated on the pole $z_I$ where $P_I(z)$ becomes on-shell. 

Now summing over the degrees of the sub-integrands $0 \leq q \leq k-1$, numbers of loops in each sub-integrand $0 \leq m \leq \ell$ and symmetrising over all internal regions, then every MHV diagram for the $\ell$-loop $\Nk$ containing the channel $I$ appears once in the expansion of the factorisation channel $I$.


\subsection*{Forward Terms}

In addition, there are forward terms in the recursion relation, arising from channels $I$ with momentum $P_I=(x-x_i)$ bounding an external and an internal region. These terms involve the forward limit of the N$^{k+1}$MHV integrand with $(l-1)$ loops and have the following contribution to the recursion relation

\begin{equation}
\label{RecursionForward}
\frac{1}{P^2_I}\int \rd^4\eta_I\; A_{n+2,k+1}^{(\ell-1)}(I,-I,i,\ldots,i-1;z_I)\, .
\end{equation}

\ni which must be symmetrised over the internal region momenta in the forward channel. All forward terms are found by summing over $i=1,\ldots,n$.

We expand the $(\ell-1)$-loop N$^{k+1}$MHV integrand using the MHV expansion. The terms involving forward limits in between particles on the same vertex vanish upon fermionic integration. The remaining terms again depend on the on-shell momentum $P_I(z_I)$ only through the holomorphic spinor $\lb_I = P_I\,|\zeta]$ and hence every term in the expansion is an MHV diagram for the original integrand. However, again all propagators bounding external regions, except for $1/P_I^2$, are shifted and evaluated on the pole $z_I$ where $P_I(z)$ becomes null. 

Once we have symmetrised over the choice of internal region in the forward channel, then each MHV diagram for the $\ell$-loop $\Nk$ integrand containing the channel $I$ appears exactly once in the expansion of the forward channel $I$.


\subsection*{Completion of Proof}

Consider now the particular MHV diagram with channels $\{I_1,\ldots,I_P\}$ where $P = (k+2\ell)$ is the number of channels in each MHV diagram for the $\ell$-loop $\Nk$ integrand. Suppose that only $P'\leq P$ of the channels are bounding external regions and are therefore affected by the all-line shift. We choose to order the channels so that those bounding external regions appear first $\{I_1,\ldots,I_{P'},\ldots,I_P\}$. 

This particular diagram occurs exactly $P'$ times in the recursion relation - once for each channel $\{I_1,\ldots,I_{P'}\}$ bounding an external region. Each of the terms corresponds to the same MHV diagram, except that $(P'-1)$ of the propagators are shifted and evaluated on the pole of the remaining one. The dependence on the shifted propagators may be factored out of each term, leaving the correct contribution from the MHV diagram, multiplied by a ratio of propagators. For example, the term arising from the channel $I_1$ contributes
\pagebreak

\begin{equation}
A^{(\ell)}_{n,k}(I_1,\ldots,I_P)\;  \frac{P_{I_2}^2 \ldots P_{I_{P'}}^2}{ P^2_{I_2}(z_{I_1}) \ldots P^2_{I_{P'}}(z_{I_1}) }\, .
\end{equation}

\ni where we have denoted the contribution from the MHV diagram with channels $\{I_1,\ldots,I_P\}$ by $A_{n,k}^{(0)}(I_1,\ldots,I_P)$. Now summing over all of the terms we find

\begin{equation}
A^{(\ell)}_{n,k}(I_1,\ldots,I_P)\;  \sum\limits_{j=1}^{P'} \frac{P_{I_1}^2 \ldots P_{I_{P'}}^2}{ P_{I_1}^2(z_{I_j}) \ldots P^2_{I_{j\!-\!1}}(z_{I_j}) P_{I_j}^2  P^2_{I_{j\!+\!1}}(z_{I_j})\ldots P_{I_{P'}}^2(z_{I_j}) }\, .
\end{equation}

\ni However, we may use the result of the contour integral

\begin{equation}
\oint \frac{\rd z}{z} \frac{P_{I_1}^2 \ldots P_{I_{P'}}^2}{ P_{I_1}^2(z) \ldots P_{I_{P'}}^2(z) } = 0
\end{equation}

\ni so that the summation collapses to unity and we recover only the contribution from the MHV diagram
\begin{equation}
 A_{n,k}^{(k)}(I_1,\ldots,I_P)
 \end{equation}
Now running over all of the allowed MHV diagrams for the $\ell$-loop $\Nk$ integrand we pick up each term in the recursion relation exactly once. Hence the recursion relation has generated the MHV diagram expansion for the $\ell$-loop N$^k$MHV integrand. This completes the induction.






\section{Conclusion and Discussion}
\label{sec:Conclusions}

\subsection*{Summary}
We have introduced a recursion relation for loop integrands in $\cN=4$ SYM generated by the all-line shift of momentum twistors
\begin{equation}
Z_i \longrightarrow Z_i + zr_i Z_*\, .
\end{equation}

\ni where the reference twistor takes the form $Z = (0,\zeta^{\dal},0)$. We have formulated both in momentum space and momentum twistor space and examined the simple examples at one and two loops. In the examples, we have found that by choosing the same reference twistor $Z_*$ at each stage of the recursion, then the solution becomes the MHV vertex expansion in both momentum space and momentum twistor space. Finally, we have proven by induction that the general solution of the recursion relation is indeed the MHV diagram expansion for all loop integrands in $\cN=4$ SYM.

\subsection*{Comparison to BCFW}
The all-line recursion relation may be compared to the BCFW recursion relation which is generated by shifting a single momentum twistor~\cite{ArkaniHamed:2010kv}
\begin{equation}
Z_1 \longrightarrow Z_1 + z Z_2\, .
\end{equation}

\ni In momentum twistor space, the relationship between terms in the BCFW expansion and MHV diagrams appear very close - choosing $Z_* = Z_1$ immediately reproduces the BCFW expressions for the tree-level NMHV amplitude and the one-loop MHV integrand (although the picture is more complex in general). In the tree-level NMHV case, this has an interpretation as a residue theorem of the grassmannian formula~\cite{ArkaniHamed:2009sx}, and it seems natural that there should be a similar remarkable story for the loop integrands too.

\subsection*{Directions for Further Research}

An clear direction for further study is to extend the all-line recursion relation beyond $\cN=4$ SYM. The forward limits appearing in the recursion relation have been studied in~\cite{CaronHuot:2010zt} and are well-defined in massless gauge theories with at least $\cN=1$ supersymmetry. In this case, the one-loop amplitudes in $\cN=1$ are cut-constructible in four dimensions, and there is evidence that the MHV formalism correctly reproduces the one-loop amplitudes~\cite{Bedford:2004py,Quigley:2004pw}. Therefore one expects that the all-line recursion relation extends immediately to massless theories with at least $\cN=1$ supersymmetry.

The proof of the MHV formalism presented in section~\ref{sec:Proof} depended on two key elements:
\begin{itemize}
\item The integrands vanish as $z\rightarrow\infty$.
\item The forward limit of neighboring legs on the same MHV vertex vanishes.
\end{itemize}
In fact, both of these properties hold for massless gauge theories with at least $\cN=1$ supersymmetry, and therefore the the proof should extend to such theories essentially unchanged. For non-supersymmetric theories, one encounters double poles in four dimensions, or alternatively, the recursion relation may be formulated in $D$-dimensions. We will return to discuss these issues in future work~\cite{FuturewithTom}

Finally, there has been interesting recent developments understanding the amplitude - Wilson loop correspondence~\cite{Mason:2010yk,Eden:2010ce,CaronHuot:2010ek}. In particular, the supersymmetric Wilson loop in momentum twistor space presented in~\cite{Mason:2010yk} generates the MHV vertex expansion when expanded perturbatively in all known examples. However, showing that the supersymmetric Wilson loop directly satisfies the all-line recursion relation in momentum twistor space would prove that it indeed computes the complete planar S-matrix.


\acknowledgments I would like to thank Timothy Adamo, Henriette Elvang, Michael Keirmaier, and especially Lionel Mason and David Skinner for useful discussions and comments on the draft. I would also like to thank the organisers of the PITP Summer School 2010 at the IAS in Princeton where I first learnt about loop-level recursion relations. I am supported by an STFC studentship.

\bibliographystyle{JHEP}

\begin{thebibliography}{10}

\bibitem{Witten:2003nn}
E.~Witten, {\it {Perturbative gauge theory as a string theory in twistor
  space}},  {\em Commun. Math. Phys.} {\bf 252} (2004) 189--258
  [\href{http://arXiv.org/abs/hep-th/0312171}{{\tt hep-th/0312171}}].

\bibitem{Cachazo:2004kj}
F.~Cachazo, P.~Svrcek and E.~Witten, {\it {MHV vertices and tree amplitudes in
  gauge theory}},  {\em JHEP} {\bf 09} (2004) 006
  [\href{http://arXiv.org/abs/hep-th/0403047}{{\tt hep-th/0403047}}].

\bibitem{Georgiou:2004wu}
G.~Georgiou and V.~V. Khoze, {\it {Tree amplitudes in gauge theory as scalar
  MHV diagrams}},  {\em JHEP} {\bf 05} (2004) 070
  [\href{http://arXiv.org/abs/hep-th/0404072}{{\tt hep-th/0404072}}].

\bibitem{Georgiou:2004by}
G.~Georgiou, E.~W.~N. Glover and V.~V. Khoze, {\it {Non-MHV Tree Amplitudes in
  Gauge Theory}},  {\em JHEP} {\bf 07} (2004) 048
  [\href{http://arXiv.org/abs/hep-th/0407027}{{\tt hep-th/0407027}}].

\bibitem{Boels:2007pj}
R.~Boels and C.~Schwinn, {\it {CSW rules for a massive scalar}},  {\em Phys.
  Lett.} {\bf B662} (2008) 80--86 [\href{http://arXiv.org/abs/0712.3409}{{\tt
  0712.3409}}].

\bibitem{Boels:2008du}
R.~Boels and C.~Schwinn, {\it {CSW rules for massive matter legs and glue
  loops}},  {\em Nucl. Phys. Proc. Suppl.} {\bf 183} (2008) 137--142
  [\href{http://arXiv.org/abs/0805.4577}{{\tt 0805.4577}}].

\bibitem{Brandhuber:2004yw}
A.~Brandhuber, B.~J. Spence and G.~Travaglini, {\it {One-loop gauge theory
  amplitudes in N = 4 super Yang-Mills from MHV vertices}},  {\em Nucl. Phys.}
  {\bf B706} (2005) 150--180 [\href{http://arXiv.org/abs/hep-th/0407214}{{\tt
  hep-th/0407214}}].

\bibitem{Bedford:2004py}
J.~Bedford, A.~Brandhuber, B.~J. Spence and G.~Travaglini, {\it {A twistor
  approach to one-loop amplitudes in N = 1 supersymmetric Yang-Mills theory}},
  {\em Nucl. Phys.} {\bf B706} (2005) 100--126
  [\href{http://arXiv.org/abs/hep-th/0410280}{{\tt hep-th/0410280}}].

\bibitem{Quigley:2004pw}
C.~Quigley and M.~Rozali, {\it {One-loop MHV amplitudes in supersymmetric gauge
  theories}},  {\em JHEP} {\bf 01} (2005) 053
  [\href{http://arXiv.org/abs/hep-th/0410278}{{\tt hep-th/0410278}}].

\bibitem{Mansfield:2005yd}
P.~Mansfield, {\it {The Lagrangian origin of MHV rules}},  {\em JHEP} {\bf 03}
  (2006) 037 [\href{http://arXiv.org/abs/hep-th/0511264}{{\tt
  hep-th/0511264}}].

\bibitem{Ettle:2006bw}
J.~H. Ettle and T.~R. Morris, {\it {Structure of the MHV-rules Lagrangian}},
  {\em JHEP} {\bf 08} (2006) 003
  [\href{http://arXiv.org/abs/hep-th/0605121}{{\tt hep-th/0605121}}].

\bibitem{Mason:2005zm}
L.~J. Mason, {\it {Twistor actions for non-self-dual fields: A derivation of
  twistor-string theory}},  {\em JHEP} {\bf 10} (2005) 009
  [\href{http://arXiv.org/abs/hep-th/0507269}{{\tt hep-th/0507269}}].

\bibitem{Boels:2006ir}
R.~Boels, L.~Mason and D.~Skinner, {\it {Supersymmetric gauge theories in
  twistor space}},  {\em JHEP} {\bf 02} (2007) 014
  [\href{http://arXiv.org/abs/hep-th/0604040}{{\tt hep-th/0604040}}].

\bibitem{Boels:2007qn}
R.~Boels, L.~Mason and D.~Skinner, {\it {From Twistor Actions to MHV
  Diagrams}},  {\em Phys. Lett.} {\bf B648} (2007) 90--96
  [\href{http://arXiv.org/abs/hep-th/0702035}{{\tt hep-th/0702035}}].

\bibitem{Bullimore:2010pj}
M.~Bullimore, L.~Mason and D.~Skinner, {\it {MHV Diagrams in Momentum Twistor
  Space}},  \href{http://arXiv.org/abs/1009.1854}{{\tt 1009.1854}}.

\bibitem{Mason:2010yk}
L.~Mason and D.~Skinner, {\it {The Complete Planar S-matrix of N=4 SYM as a
  Wilson Loop in Twistor Space}},  \href{http://arXiv.org/abs/1009.2225}{{\tt
  1009.2225}}.

\bibitem{Britto:2004ap}
R.~Britto, F.~Cachazo and B.~Feng, {\it {New Recursion Relations for Tree
  Amplitudes of Gluons}},  {\em Nucl. Phys.} {\bf B715} (2005) 499--522
  [\href{http://arXiv.org/abs/hep-th/0412308}{{\tt hep-th/0412308}}].

\bibitem{Britto:2005fq}
R.~Britto, F.~Cachazo, B.~Feng and E.~Witten, {\it {Direct Proof Of Tree-Level
  Recursion Relation In Yang- Mills Theory}},  {\em Phys. Rev. Lett.} {\bf 94}
  (2005) 181602 [\href{http://arXiv.org/abs/hep-th/0501052}{{\tt
  hep-th/0501052}}].

\bibitem{Bern:2005hs}
Z.~Bern, L.~J. Dixon and D.~A. Kosower, {\it {On-shell recurrence relations for
  one-loop QCD amplitudes}},  {\em Phys. Rev.} {\bf D71} (2005) 105013
  [\href{http://arXiv.org/abs/hep-th/0501240}{{\tt hep-th/0501240}}].

\bibitem{Bern:2005cq}
Z.~Bern, L.~J. Dixon and D.~A. Kosower, {\it {Bootstrapping multi-parton loop
  amplitudes in QCD}},  {\em Phys. Rev.} {\bf D73} (2006) 065013
  [\href{http://arXiv.org/abs/hep-ph/0507005}{{\tt hep-ph/0507005}}].

\bibitem{ArkaniHamed:2010kv}
N.~Arkani-Hamed, J.~L. Bourjaily, F.~Cachazo, S.~Caron-Huot and J.~Trnka, {\it
  {The All-Loop Integrand For Scattering Amplitudes in Planar N=4 SYM}},
  \href{http://arXiv.org/abs/1008.2958}{{\tt 1008.2958}}.

\bibitem{Risager:2005vk}
K.~Risager, {\it {A direct proof of the CSW rules}},  {\em JHEP} {\bf 12}
  (2005) 003 [\href{http://arXiv.org/abs/hep-th/0508206}{{\tt
  hep-th/0508206}}].

\bibitem{Elvang:2008vz}
H.~Elvang, D.~Z. Freedman and M.~Kiermaier, {\it {Proof of the MHV vertex
  expansion for all tree amplitudes in N=4 SYM theory}},  {\em JHEP} {\bf 06}
  (2009) 068 [\href{http://arXiv.org/abs/0811.3624}{{\tt 0811.3624}}].

\bibitem{Cohen:2010mi}
T.~Cohen, H.~Elvang and M.~Kiermaier, {\it {On-shell constructibility of tree
  amplitudes in general field theories}},
  \href{http://arXiv.org/abs/1010.0257}{{\tt 1010.0257}}.

\bibitem{Boels:2010nw}
R.~H. Boels, {\it {On BCFW shifts of integrands and integrals}},
  \href{http://arXiv.org/abs/1008.3101}{{\tt 1008.3101}}.

\bibitem{Drummond:2009fd}
J.~M. Drummond, J.~M. Henn and J.~Plefka, {\it {Yangian symmetry of scattering
  amplitudes in N=4 super Yang-Mills theory}},  {\em JHEP} {\bf 05} (2009) 046
  [\href{http://arXiv.org/abs/0902.2987}{{\tt 0902.2987}}].

\bibitem{Hodges:2009hk}
A.~Hodges, {\it {Eliminating spurious poles from gauge-theoretic amplitudes}},
  \href{http://arXiv.org/abs/0905.1473}{{\tt 0905.1473}}.

\bibitem{Mason:2009qx}
L.~Mason and D.~Skinner, {\it {Dual Superconformal Invariance, Momentum
  Twistors and Grassmannians}},  {\em JHEP} {\bf 11} (2009) 045
  [\href{http://arXiv.org/abs/0909.0250}{{\tt 0909.0250}}].

\bibitem{Hodges:2010kq}
A.~Hodges, {\it {The box integrals in momentum-twistor geometry}},
  \href{http://arXiv.org/abs/1004.3323}{{\tt 1004.3323}}.

\bibitem{Mason:2010pg}
L.~Mason and D.~Skinner, {\it {Amplitudes at Weak Coupling as Polytopes in
  AdS5}},  \href{http://arXiv.org/abs/1004.3498}{{\tt 1004.3498}}.

\bibitem{Drummond:2010mb}
J.~M. Drummond and J.~M. Henn, {\it {Simple loop integrals and amplitudes in
  N=4 SYM}},  \href{http://arXiv.org/abs/1008.2965}{{\tt 1008.2965}}.

\bibitem{Alday:2010jz}
L.~F. Alday, {\it {Some analytic results for two-loop scattering amplitudes}},
  \href{http://arXiv.org/abs/1009.1110}{{\tt 1009.1110}}.

\bibitem{Drummond:2008vq}
J.~M. Drummond, J.~Henn, G.~P. Korchemsky and E.~Sokatchev, {\it {Dual
  superconformal symmetry of scattering amplitudes in N=4 super-Yang-Mills
  theory}},  {\em Nucl. Phys.} {\bf B828} (2010) 317--374
  [\href{http://arXiv.org/abs/0807.1095}{{\tt 0807.1095}}].

\bibitem{Sever:2009aa}
A.~Sever and P.~Vieira, {\it {Symmetries of the N=4 SYM S-matrix}},
  \href{http://arXiv.org/abs/0908.2437}{{\tt 0908.2437}}.

\bibitem{Brandhuber:2010mi}
A.~Brandhuber, B.~Spence, G.~Travaglini and G.~Yang, {\it {A Note on Dual MHV
  Diagrams in N=4 SYM}},  \href{http://arXiv.org/abs/1010.1498}{{\tt
  1010.1498}}.

\bibitem{Drummond:2008cr}
J.~M. Drummond and J.~M. Henn, {\it {All tree-level amplitudes in N=4 SYM}},
  {\em JHEP} {\bf 04} (2009) 018 [\href{http://arXiv.org/abs/0808.2475}{{\tt
  0808.2475}}].

\bibitem{Bullimore:2010pa}
M.~Bullimore, {\it {Inverse Soft Factors and Grassmannian Residues}},
  \href{http://arXiv.org/abs/1008.3110}{{\tt 1008.3110}}.

\bibitem{ArkaniHamed:2009sx}
N.~Arkani-Hamed, J.~Bourjaily, F.~Cachazo and J.~Trnka, {\it {Local Spacetime
  Physics from the Grassmannian}},  \href{http://arXiv.org/abs/0912.3249}{{\tt
  0912.3249}}.

\bibitem{CaronHuot:2010zt}
S.~Caron-Huot, {\it {Loops and trees}},
  \href{http://arXiv.org/abs/1007.3224}{{\tt 1007.3224}}.

\bibitem{FuturewithTom}
M.~Bullimore and T.~Melia, {\it {Work in Progress}}, .

\bibitem{Eden:2010ce}
B.~Eden, G.~P. Korchemsky and E.~Sokatchev, {\it {More on the duality
  correlators/amplitudes}},  \href{http://arXiv.org/abs/1009.2488}{{\tt
  1009.2488}}.

\bibitem{CaronHuot:2010ek}
S.~Caron-Huot, {\it {Notes on the scattering amplitude / Wilson loop duality}},
   \href{http://arXiv.org/abs/1010.1167}{{\tt 1010.1167}}.

\end{thebibliography}

\end{document}